\documentclass[preprint,5p,times,twocolumn]{elsarticle}

\usepackage[utf8]{inputenc}
\usepackage{bm}
\usepackage{amsmath,amssymb}
\usepackage{graphicx}
\usepackage{epstopdf}
\usepackage{wrapfig}
\usepackage{array}
\usepackage{listings}
\usepackage[para,online,flushleft]{threeparttablex}
\usepackage{booktabs,dcolumn}
\usepackage{color}
\usepackage{textpos}
\usepackage{booktabs}
\usepackage{multirow,bigdelim}
\usepackage{float}
\usepackage{upgreek} 
\usepackage{hyperref}
\usepackage{xcolor}
\usepackage{comment}
\usepackage{physics}
\usepackage{braket}
\usepackage{environ}
\usepackage{bbm}
\usepackage{xspace}
\usepackage{isotope}

\newcolumntype{C}[1]{>{\centering\arraybackslash}p{#1}}

\hypersetup{breaklinks=true,colorlinks=true,linkcolor=blue,citecolor=blue,
filecolor=magenta,urlcolor=blue}

\makeatletter
\def\@bibdataout@aps{%
\immediate\write\@bibdataout{%
@CONTROL{%
apsrev41Control%
\longbibliography@sw{%
    ,author="08",editor="1",pages="1",title="0",year="1"%
    }{%
    ,author="08",editor="1",pages="1",title="",year="1"%
    }%
  }%
}%
\if@filesw \immediate \write \@auxout {\string \citation {apsrev41Control}}\fi
}
\makeatother

\NewEnviron{subalign}[1][]{%
\begin{subequations}\begin{align}
  \BODY
\end{align}\label{#1}\end{subequations}
}

\NewEnviron{spliteq}{%
\begin{equation}\begin{split}
  \BODY
\end{split}\end{equation}
}

\journal{Physics Letters B}

\newcommand{\beginsupplement}{%
	\setcounter{table}{0}
	\renewcommand{\thetable}{S\arabic{table}}%
	\setcounter{figure}{0}
	\renewcommand{\thefigure}{S\arabic{figure}}%
	\setcounter{equation}{0}
	\renewcommand{\theequation}{S\arabic{equation}}        
}

\begin{document}
\begin{frontmatter}

\title{\emph{Ab Initio} Calculations of the Carbon and Oxygen Isotopes:\\ Energies, Correlations, and Superfluid Pairing}

\author[IRIS]{Young-Ho Song}
\address[IRIS]{Institute for Rare Isotope Sciences, Institute for Basic Science, Daejeon 34000, Korea}

\author[CENS]{Myungkuk Kim}
\address[CENS]{Center for Exotic Nuclear Studies, Institute for Basic Science, Daejeon 34126, Korea}

\author[CENS]{Youngman Kim}

\author[KISTI]{Kihyeon Cho}
\address[KISTI]{Korea Institute of Science and Technology Information, Daejeon 34141, Korea}

\author[Gaziantep,KFUPM]{Serdar Elhatisari}
\address[Gaziantep]{Faculty of Natural Sciences and Engineering, Gaziantep Islam Science and Technology University, Gaziantep 27010, Turkey}
\address[KFUPM]{King Fahd University of Petroleum and Minerals (KFUPM), 31261 Dhahran, Saudi Arabia}

\author[FRIB]{Dean Lee}
\address[FRIB]{Facility for Rare Isotope Beams and Department of Physics and Astronomy, Michigan State University, MI 48824, USA}

\author[FRIB]{Yuan-Zhuo Ma}

\author[HI,IAS,Peng]{Ulf-G.~Meißner}
\address[HI]{Helmholtz-Institut f\"ur Strahlen- und Kernphysik (Theorie) and Bethe Center for Theoretical Physics, Universitat Bonn, D-53115 Bonn, Germany}
\address[IAS]{Institute for Advanced Simulation (IAS-4),  D-52425 J\"ulich, Germany}
\address[Peng]{Peng Huanwu Collaborative Center for Research and Education, International Institute for Interdisciplinary and Frontiers, Beihang University, Beijing 100191, China}

\begin{abstract}
We perform \textit{ab initio} nuclear lattice calculations of the neutron-rich carbon and oxygen isotopes using high-fidelity chiral interactions.  We find good agreement with the observed binding energies and compute correlations associated with each two-nucleon interaction channel.  For the isospin $T=1$ channels, we show that the dependence on $T_z$ provides a measure of the correlations among the extra neutrons in the neutron-rich nuclei.  For the spin-singlet S-wave channel, we observe that any paired neutron interacts with the nuclear core as well as its neutron pair partner, while any unpaired neutron interacts primarily with only the nuclear core.  For the other partial waves, the correlations among the extra neutrons grow more slowly and smoothly with the number of neutrons.  These general patterns are observed in both the carbon and oxygen isotopes and may be universal features that appear in many neutron-rich nuclei.
\end{abstract}

\begin{keyword}
Nuclear Lattice Effective Field Theory\sep Neutron-rich Isotopes \sep Superfluid parings \sep Ab Initio calculation
\end{keyword}

\end{frontmatter}

Nuclei far from the valley of stability provide a valuable laboratory for probing the dependence on nuclear forces and the nature of the quantum correlations among nucleons.  There have been several \textit{ab initio} calculations of neutron-rich oxygen isotopes \cite{Hagen:2009mm,Otsuka:2009cs,Hagen:2012sh, Holt:2012fr, Cipollone:2013zma, Jansen:2014qxa, Stroberg:2019bch,Ma:2020spz,Kaur:2022yoh} as well as neutron-rich carbon isotopes \cite{Forssen:2011dr,Jansen:2014qxa,Kanungo:2016tmz,Tran:2017gxv,Stroberg:2019bch,Li:2024kno}.   In this work, we perform calculations of neutron-rich carbon and oxygen isotopes using nuclear lattice effective field theory (NLEFT).  We use chiral effective field theory (EFT) interactions defined on a three-dimensional lattice and perform quantum Monte Carlo simulations of the many-body system using auxiliary fields.  Reviews of NLEFT and related methods can be found in Refs.~\cite{Lee:2008fa,Drut:2012md,Lee:2016fhn,Lahde:2019npb}, and reviews of chiral EFT can be found in Refs.~\cite{Epelbaum:2008ga,Machleidt:2016rvv, Hammer:2019poc}.

Wavefunction matching was introduced in Ref.~\cite{Elhatisari:2022zrb} to accelerate the convergence of perturbation theory.  We also use wavefunction matching in this work and apply the interactions defined in Ref.~\cite{Elhatisari:2022zrb} with spatial lattice spacing $a = 1.32$~fm.  Details of the interactions and computational methods can be found in the Supplemental Material accompanying Ref.~\cite{Elhatisari:2022zrb}. For our chiral interactions, a low-energy scheme is used where the two-nucleon two-pion exchange and higher-pion exchange interactions are treated as short-range contact interactions.  Within this framework, we include all two-nucleon and three-nucleon interactions up to $O(Q^4)$ or next-to-next-to-next-to-leading order (N3LO).  This includes chiral three-nucleon interactions such as the one-pion exchange, two-pion exchange, and short-range three-nucleon interactions.  As introduced in Ref.~\cite{Elhatisari:2022zrb}, we also include additional three-nucleon interactions that correspond with specific choices for the local regulators used in the three-nucleon interactions.  We have not included any four-nucleon interactions.

\begin{figure}[h]
\centering 
\includegraphics[width=9cm]{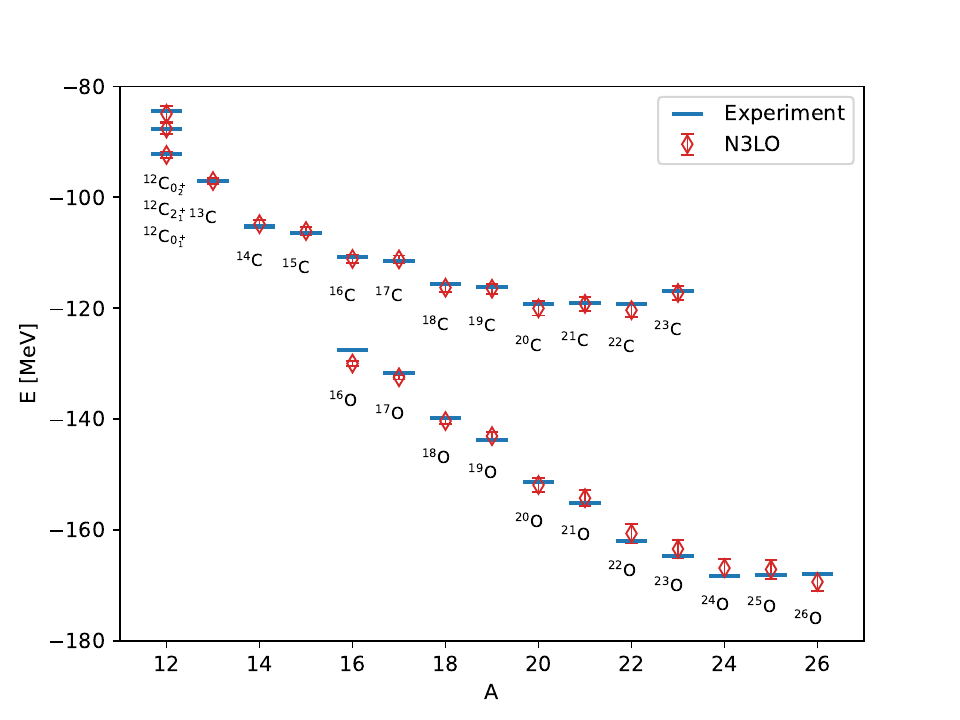}
\caption{Ground state energies for the neutron-rich carbon and oxygen isotopes.  NLEFT results at order N3LO are compared with experimental data. In the case of $^{12}$C, we also show the first two excited states.
Experimental data are from \cite{Kelley:2017qgh} \cite{Wang:2021xhn}.
}
\label{fig:energies}
\end{figure}

In Fig.~\ref{fig:energies}, we present lattice results for the energies of the neutron-rich carbon and oxygen isotopes versus the number of nucleons, $A$.  The energies for $^{12-14}$C, of the first two excited states in $^{12}$C and $^{16-18}$O were already reported in Ref.~\cite{Elhatisari:2022zrb}, and they are shown again in the results here.  The error bars correspond to one standard deviation and include statistical errors as well as uncertainties in the extrapolation to infinite Euclidean time and infinite volume.  While there are some small deviations in comparison with experimental data, the overall agreement is quite good.  In future work, we plan to investigate the remaining sources of errors and perform calculations of other observables such as charge radii, quadrupole moments, electromagnetic transitions, and magnetic dipole moments.

Having demonstrated that the lattice calculations accurately reproduce the energies of the neutron-rich carbon and oxygen isotopes, we now turn our attention to probing the dependence on nuclear forces and measuring quantum correlations.  In each partial-wave channel, we calculate $\braket{\Psi|\Delta O|\Psi}$ for some perturbing two-nucleon operator $\Delta O$. 
Similar sensitivity studies have been performed in the literature \cite{Ekstrom:2019lss,Hu:2021trw}. 
In our analysis, however, we do not focus on the details of $\Delta O$ but rather the change to the scattering phase shifts, $\Delta \delta(p)$.  By relying on physical observables, we are constructing a model-independent framework that can be translated to any low-energy EFT calculation. 
We mean `model-independent’ in the sense that one can compare two EFTs
regardless of their details in operator structure by
matching low-energy observables. Two different EFT calculations would simply agree on $\Delta \delta(p)$ and determine their corresponding operators $\Delta O$ accordingly.  Induced higher-body operators can also be determined by matching to higher-body physical observables. 

For each partial-wave channel, we consider a short-range two-nucleon interaction operator that, when added to the full Hamiltonian, produces a $1\%$ reduction in the scattering phase shift at relative momentum $p=150$~MeV.  The detailed form of the operators we use and their effect on the scattering phase shifts are described in the Supplemental Material~\cite{SM}.  Before presenting lattice results for the two-nucleon correlations, we first prove a useful fact about isospin correlations that we call $T_z$ linearity.  More specifically, we will consider the `energy correlation' between two nucleons.

Let $\ket{\Psi_{(1/2,-1/2)}}$ be a nuclear state with isospin $T=1/2$ and $T_z=-1/2$.  For example, $\ket{\Psi_{(1/2,-1/2)}}$ could be the ground of a nucleus such as $^{13}$C or $^{17}$O with one more neutron than the number of protons. Let $A_{(1,T_z)}$ be an operator with isospin $T=1$ and arbitrary $T_z$.  For example, $A_{(1,T_z)}$ could be a short-range operator that annihilates two nucleons in some $T=1$ partial-wave channel.  Then $T_z=-1$ corresponds to the annihilation of two protons, $T_z=0$ corresponds to the isospin-symmetric annihilation of a proton and neutron, and $T_z=1$ corresponds to the annihilation of two neutrons.  We now consider the operator expectation value, 
\begin{equation}
f(T_z) = \braket{\Psi_{(1/2,-1/2)}^{}|A^\dagger_{(1,T_z)} A_{(1,T_z)}|\Psi_{(1/2,-1/2)}^{}}.
\end{equation}
We note that $A_{(1,T_z)}\ket{\Psi_{(1/2,-1/2)}}$ can be decomposed into two irreducible isospin representations, $T=3/2$ and $T=1/2$. Let us write $f_{3/2}$ for the $3/2$ amplitude and $f_{1/2}$ for the $1/2$ amplitude.  It is straightforward to show that $f(-1) = f_{3/2}$, $f(0)=\frac{2}{3}f_{3/2}+\frac{1}{3}f_{1/2}$, and $f(1)=\frac{1}{3}f_{3/2}+\frac{2}{3}f_{1/2}$. Therefore, the dependence on $T_z$ is linear, and we have the relation $f(1)=2f(0)-f(-1)$.  

Let us now consider a neutron-rich nucleus that has more than one extra neutron so that its isospin is greater than $1/2$.  We can still define $f(T_z)$ in the same manner, 
\begin{equation}
f(T_z) = \braket{\Psi|A^\dagger_{(1,T_z)} A_{(1,T_z)}|\Psi}.
\end{equation}
We now compare $f(1)$ against the linear combination $2f(0)-f(-1)$.  If each of the extra neutrons are uncorrelated with each other, then the additional correlations produced by each extra neutron are additive, and we expect $T_z$ linearity to still hold, $f(1)=2f(0)-f(-1)$.  In general, however, there will be some correlations among the extra neutrons, and this results in $f(1)$ being different from $2f(0)-f(-1)$.  The comparison between $f(1)$ and $2f(0)-f(-1)$ is therefore a measure of correlations among the extra neutrons in a neutron-rich nucleus.

\begin{figure}[h]
\centering 
\includegraphics[width=8cm]
{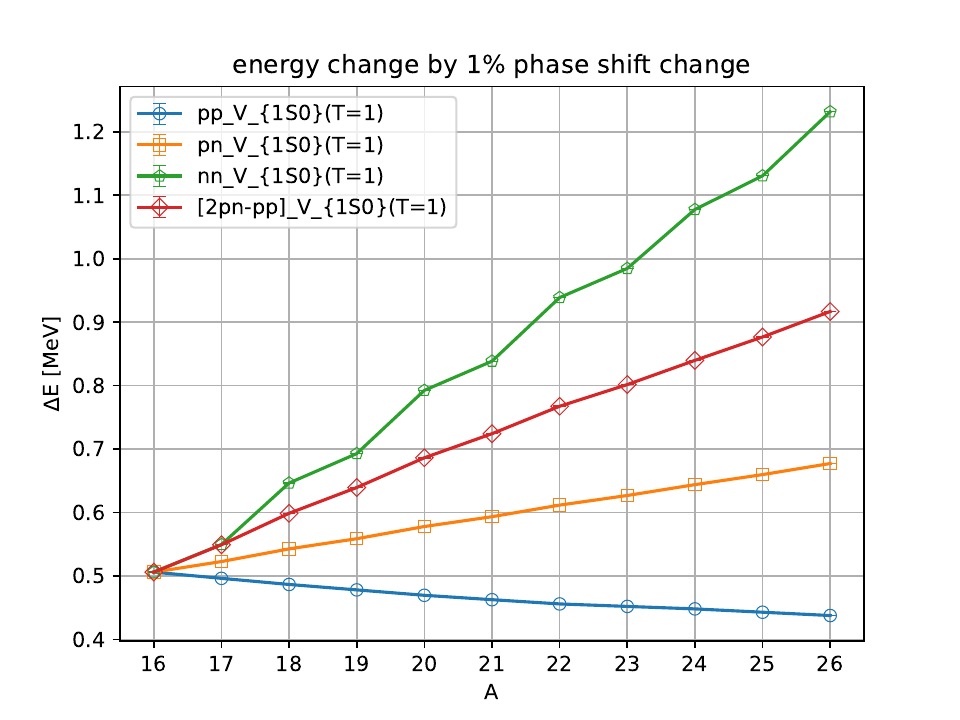} 
\includegraphics[width=8cm]
{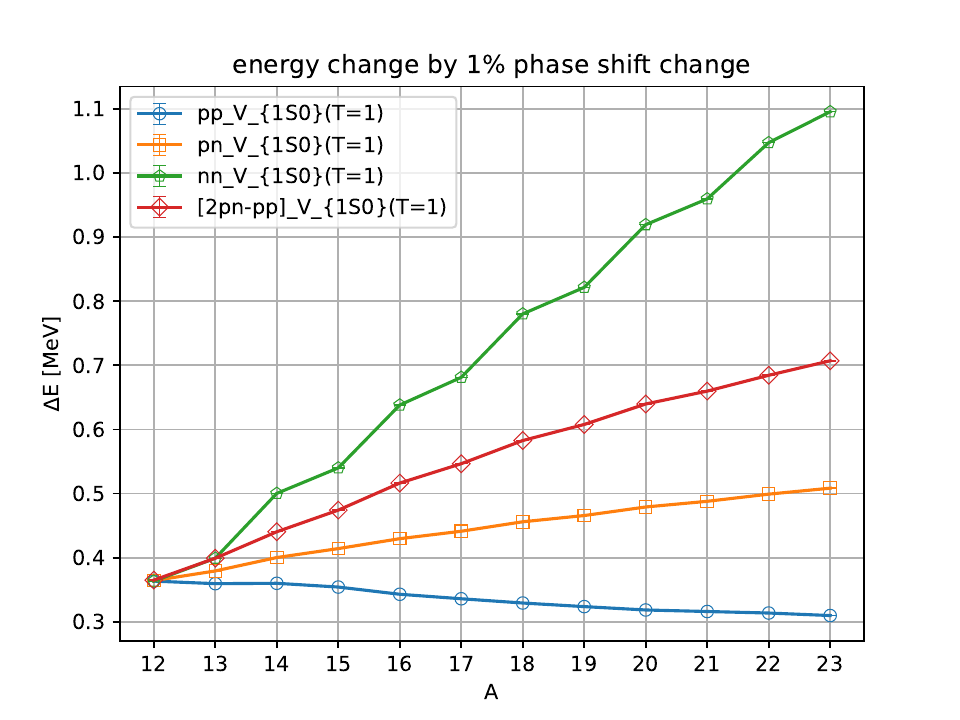} 
\caption{Correlations for pp, pn, nn, and 2pn$-$pp in the $^1$S$_0$ channel.  The top panel shows the oxygen isotopes, and the bottom panel shows the carbon isotopes.}
\label{fig:1S0}
\end{figure}

In Fig.~\ref{fig:1S0}, we show $^1$S$_0$ correlations for the combinations 
 proton-proton (pp), proton-neutron (pn), neutron-neutron (nn), and twice proton-neutron minus proton-proton (2pn$-$pp).  The top panel shows the oxygen isotopes, and the bottom panel shows the carbon isotopes.  In both cases, the pp correlations are rather flat, decreasing by only $14\%$ from $^{16}$O to $^{26}$O and decreasing only $15\%$ from $^{12}$C to $^{23}$C.  This is an indication that the proton structure of the nuclear core does not change much. Previous lattice simulations have shown that the ground states of $^{16}$O and $^{12}$C both have significant alpha cluster substructures \cite{Epelbaum:2011md,Epelbaum:2012qn,Epelbaum:2013paa,Freer:2017gip,Summerfield:2021oex,Shen:2021kqr,Shen:2022bak}.  Our results here suggest that the pp correlations within the alpha clusters remain mostly intact as extra neutrons are added.

We see that the $^1$S$_0$ nn correlations for oxygen and carbon both have a prominent ``staircase'' pattern produced by superfluid pairing.  We note that the pp, pn, nn correlations are equal for $^{16}$O and for $^{12}$C due to isospin symmetry.  Due to $T_z$ linearity, we observe that the nn correlations equal the 2pn$-$pp correlations for $^{17}$O and for $^{13}$C.  In each of the correlation measurements presented here, we have not included perturbative theory corrections in nonperturbative wave function to the correlations.  Therefore, the correlations being measured are those associated with the nonperturbative Hamiltonian used in the propagation of the wavefunction, and the nonperturbative Hamiltonian used has exact isospin symmetry.

If we look closely at the $^1$S$_0$ nn correlations for oxygen and carbon, we see that adding an unpaired or odd neutron produces an increase in $\Delta E$ whose slope is very close to 
that of 2pn$-$pp.  See, for example, the increase from $^{18}$O to $^{19}$O, $^{20}$O to $^{21}$O, $^{14}$C to $^{15}$C, or $^{16}$C to $^{17}$C.  A simple interpretation of this result is that the unpaired neutron is only weakly correlated with the other extra neutrons and is predominantly interacting with the $T=0$ nuclear core. On the other hand, adding one more neutron to complete the $^1$S$_0$ pair produces an increase in $\Delta E$ with slope rising higher than that of 2pn$-$pp.  This additional neutron is interacting strongly with its pair partner as well as with the nuclear core.  We note that the pn correlations follow a smooth and almost linear trajectory as a function of the number of neutrons.  See Ref.~\cite{Brown:2013} for a discussion of pairing for these nuclei in the shell model.

\begin{figure}[ht]
\centering 
\includegraphics[width=8cm]
{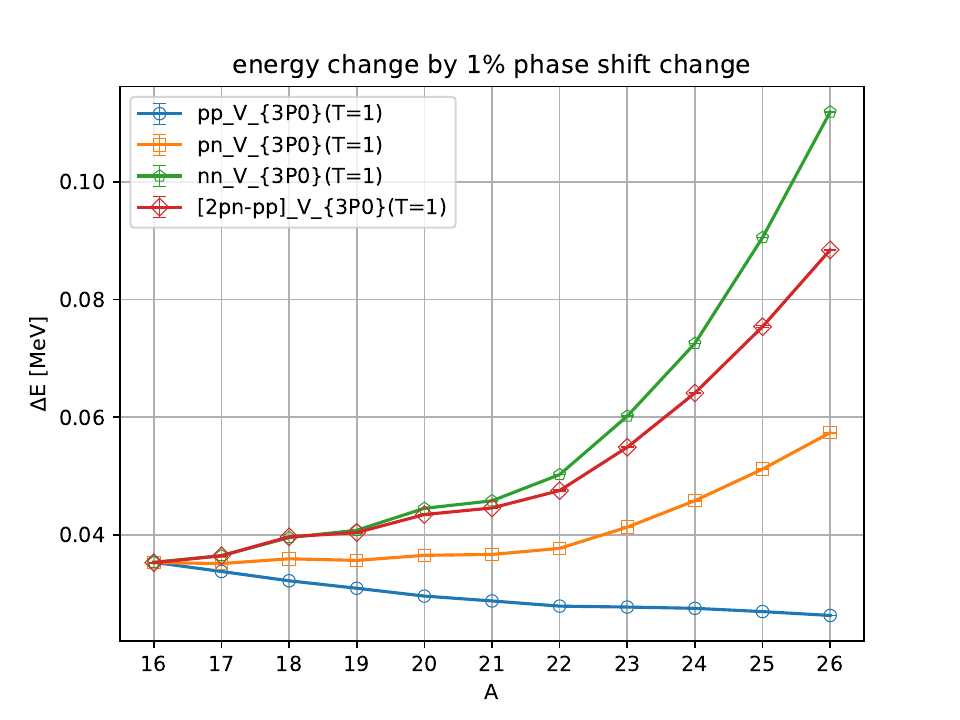} 
\includegraphics[width=8cm]
{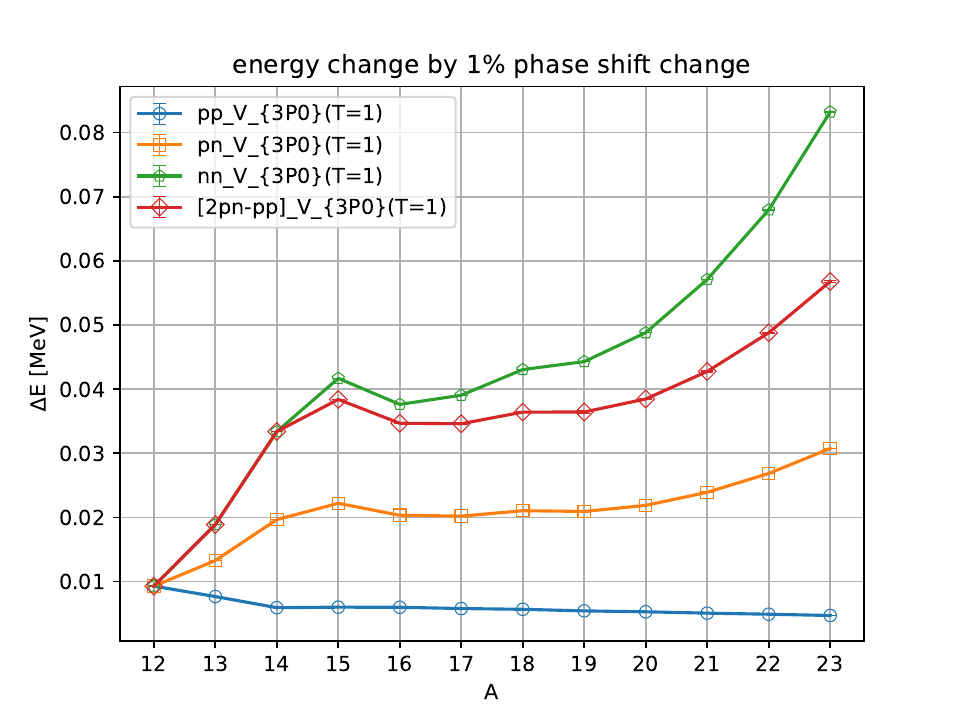} 
\caption{Correlations for pp, pn, nn, and 2pn$-$pp in the $^3$P$_0$ channel.  The top panel shows the oxygen isotopes, and the bottom panel shows the carbon isotopes.}
\label{fig:3P0}
\end{figure}

In  Fig.~\ref{fig:3P0}, we show $^3$P$_0$ correlations for pp, pn, nn, and 2pn$-$pp.  The top panel shows the oxygen isotopes, and the bottom panel shows the carbon isotopes. 
We again note that the pp, pn, nn correlations are equal for $^{16}$O and $^{12}$C due to isospin symmetry, and the nn and 2pn$-$pp correlations are equal for $^{17}$O and $^{13}$C due to $T_z$ linearity.  We observe that the $^3$P$_0$ pp correlations decrease gradually with the number of neutrons, but at a faster rate than we observed for the $^1$S$_0$ channel.  The decrease is $25\%$ from $^{16}$O to $^{26}$O, and the decrease is $49\%$ from $^{12}$C to $^{23}$C.  We note that P-wave correlations between protons would not come from protons within one alpha cluster, but rather protons from two different neighboring alpha clusters.  These results suggest that while the alpha clusters may remain intact, they may become less correlated with each other as extra neutrons are added. 

For the oxygen isotopes, we see a plateau in the $^3$P$_0$ pn correlations for $^{17}$O through $^{22}$O and then an upward slope thereafter.  This is consistent with the closure of the $1d_{5/2}$ subshell at $N=14$.  A similar plateau can be seen also in the carbon isotopes, however the situation is more complicated due to the lack of a closed proton shell and significant deformation in the proton distribution.  We see some interesting behavior in the pn and nn correlations at $^{14}$C, $^{15}$C, and $^{16}$C, which may indicate some changes to the orbital structure of the extra neutrons in the carbon isotopes. 

The nn correlations for the oxygen isotopes remain very close to the 2pn$-$pp correlations even for up to six extra neutrons.  The same is true for the carbon isotopes for up to four extra neutrons.  The $^3$P$_0$ correlations between the extra neutrons grow slowly and smoothly with the number of neutrons.  The same is true for the other $T=1$ partial waves.  We note that there are some faint oscillations in the P-wave correlations due to the pairing driven by the $^1$S$_0$ interactions.  In the Supplemental Material~\cite{SM}, we present results for the other partial waves, including both $T=1$ and $T=0$ channels.  

There has been considerable discussion in the recent literature about short-range correlations and $T=0$ proton-neutron pairs \cite{Hen:2014nza,CLAS:2018yvt,CLAS:2019vsb,Cruz-Torres:2019fum,Tropiano:2021qgf}.  These short-range correlations arise from the singular tensor force and depend strongly on the short-distance resolution scale.  In our calculations, we have used a relatively low resolution scale associated with our $1.32$~fm lattice spacing, and the total $T=0$ S-wave correlations are larger than the total $T=1$ S-wave correlations by only $26\%$ for $^{16}$O and only $25\%$ for $^{12}$C.  The near equality of the $T=0$ and $T=1$ contributions is related to the hidden spin-isospin exchange symmetry discussed in Ref.~\cite{Lee:2020esp}.

We have presented {\it ab initio} lattice results for the neutron-rich carbon and oxygen isotopes using high-fidelity chiral interactions.  The energies are in good agreement with experimental data.  We have also computed correlations associated with two-nucleon interaction operators in various partial-wave channels.  By studying the dependence on $T_z$ in the $T=1$ channels, we are able to measure correlations among the extra neutrons in the neutron-rich carbon and oxygen isotopes.  For the $^1$S$_0$ channel, we find that any paired neutron interacts with the nuclear core and its neutron pair partner, while any unpaired neutron interacts primarily with only the nuclear core.  For the other partial waves, the correlations among the extra neutrons grow slowly and smoothly with the number of neutrons.  These findings for the carbon and oxygen isotopes may in fact be universal properties that can be seen in many other neutron-rich nuclei. 

The observed ``staircase'' pattern for the $^1$S$_0$ nn correlations may have an impact on the charge radii for the carbon and oxygen isotopes with even and odd numbers of neutrons.  We plan to investigate these effects in the future using the pinhole algorithm \cite{Elhatisari:2017eno}, see
also the recent work in~\cite{Ren:2025vpe}.
However, pinhole calculations of $A$-body density correlations do not have an immediate analog for other nuclear many-body methods.  It is therefore valuable that a significant amount of information about nuclear forces and quantum correlations can be deduced from the simple correlation measurements introduced here and can be expressed in a model-independent language.  The correlation studies presented here can be readily adopted by other groups using different nuclear many-body methods.

\section*{ACKNOWLEDGMENTS}
We are grateful to many members of the NLEFT Collaboration for useful and stimulating discussions and for their important contributions in helping to develop the theoretical framework, algorithms, and computational codes needed to perform this work.  This work was supported by the Rare Isotope Science Project of Institute for Basic Science;  Ministry of Science and ICT (MSICT); National Research Foundation of Korea (2013M7A1A1075764); Institute for Basic Science (IBS-I001-01, IBS-R031-D1); the major institutional R$\&$D program, KISTI (No. K-24-L02-C04-S01, K25L2M2C3); U.S. Department of Energy (grants DE-SC0013365, DE-SC0023175, DE-SC0023658, DE-SC0024586); U.S. National Science Foundation (grant PHY-2310620); European Research Council (ERC) under the European Union’s Horizon 2020 research and innovation programme (ERC AdG EXOTIC, grant agreement No. 101018170);  CAS President’s International Fellowship Initiative (PIFI) (Grant No. 2025PD0022); Scientific and Technological Research Council of Turkey (TUBITAK project no. 123F464).  Computational resources
were provided by the National Supercomputing Center of Korea with supercomputing resources including technical support (KSC-2022-CHA-0003, KSC-2023-CRE-0006, KSC-2024-CRE-0256, KSC-2023-CHA-0005, KSC-2024-CHA-0001, KSC-2025-CHA-0004); the Gauss Centre for Supercomputing e.V. (www.gauss-centre.eu) for computing time on the GCS Supercomputer JUWELS at J{\"u}lich Supercomputing Centre (JSC) and special GPU time allocated on JURECA-DC; and the Oak Ridge Leadership Computing Facility through the INCITE award ``Ab-initio nuclear structure and nuclear reactions.''

\bibliography{References}
\bibliographystyle{elsarticle-num}

\beginsupplement

\clearpage     

\onecolumn       

\section{Supplemental Material}

\subsection{Two-Nucleon Correlation Operators}
In Ref.~\cite{Li:2018ymw}, lattice chiral interactions were developed based on partial-wave projections and nonlocal smearing functions.  For our calculations of the two-nucleon correlations, we use this method to define the two-nucleon operators. The angular dependence of the relative separation between the two nucleons is prescribed by spherical harmonics, and the dependence on the nucleon spins is given by spin-orbit Clebsch-Gordan coefficients.  We define the operators $a^{s_{\rm \rm NL}}_{\rm i,j}({\bf n})$ and $a^{s_{\rm \rm NL}\dagger}_{\rm i,j}({\bf n})$ with nonlocal smearing parameter $s_{\rm \rm NL}$, spin $i=0,1$ (up, down) and isospin $j=0,1$ (proton,neutron) indices, 
\begin{equation}
	a^{s_{\rm \rm NL}}_{\rm i,j}({\bf n})=a_{\rm i,j}({\bf n})+s_{\rm \rm NL}\sum_{\rm |{\bf n'}|=1}a_{\rm i,j}({\bf
		n}+{\bf n'}).
\end{equation}
\begin{equation}
	a^{s_{\rm \rm NL}\dagger}_{\rm i,j}({\bf n})=a^{\dagger}_{\rm i,j}({\bf n})+s_{\rm \rm NL}\sum_{\rm |{\bf
			n'}|=1}a^{\dagger}_{\rm i,j}({\bf
		n}+{\bf n'}).
\end{equation}
The nonlocal smearing can be extended beyond nearest neighbors in a straightforward manner.  We define the following two-by-two matrices to make a spin-$0$ combination,
\begin{equation}
	M_{\rm ii'}(0,0) = \frac{1}{\sqrt{2}}[\delta_{\rm i,0}\delta_{\rm i',1}-\delta_{\rm i,1}\delta_{\rm i',0}],
\end{equation}
and spin-$1$ combinations,
\begin{align}
	& M_{\rm ii'}(1,1) = \delta_{\rm i,0}\delta_{\rm i',0}, \nonumber \\
	& M_{\rm ii'}(1,0) = \frac{1}{\sqrt{2}}[\delta_{\rm i,0}\delta_{\rm i',1}+\delta_{\rm i,1}\delta_{\rm i',0}], \nonumber \\
	& M_{\rm ii'}(1,-1) = \delta_{\rm i,1}\delta_{\rm i',1}.
\end{align}

We can define the pair annihilation operators $[a({\bf n})a({\bf n'})]^{s_{\rm \rm
		NL}}_{S,S_z,T,T_z}$, where
\begin{equation}
	[a({\bf n})a({\bf n'})]^{s_{\rm \rm
			NL}}_{S,S_z,T,T_z}=\sum_{\rm i,j,i',j'} a^{s_{\rm \rm NL}}_{\rm i,j}({\bf n})M_{\rm ii'}(S,S_z)M_{\rm jj'}(T,T_z)a^{s_{\rm \rm
			NL}}_{\rm i',j'}({\bf n'}),
	\label{spin-isospin}
\end{equation}
with spin quantum numbers $S,S_z$ and isospin quantum numbers $T,T_z$.
We also define the solid harmonics
\begin{equation}
	R_{\rm L,L_z}({\bf r}) = \sqrt{\frac{4\pi}{2L+1}}r^L Y_{\rm L,L_z}(\theta,\phi),
\end{equation}
and their complex conjugates
\begin{equation}
	R^*_{\rm L,L_z}({\bf r}) = \sqrt{\frac{4\pi}{2L+1}}r^L Y^*_{\rm L,L_z}(\theta,\phi).
\end{equation}
We note that $R_{\rm L,L_z}$ and $R^*_{\rm L,L_z}$ are homogeneous polynomials with degree $L$. 

We define finite lattice difference operators ${\bf \nabla}_{\rm 1/2} $ and ${\bf \nabla}$ as follows,

\begin{equation}
	\nabla_l f({\vec{\bm n}}) = \frac{1}{2}f({\vec{\bm n}}+\hat{{\bm l}}) -\frac{1}{2}f({\vec{\bm n}}-\hat{{\bm l}})  \end{equation}
\begin{equation}
	\nabla_{1/2,l} f({\vec{\bm n}}) = f({\vec{\bm n}}+\frac{1}{2}\hat{{\bm l}}) -f({\vec{\bm n}}-\frac{1}{2}\hat{{\bm l}})  \end{equation}
Using the pair annihilation operators, lattice finite differences,
and solid harmonics, we form the operator combinations
\begin{equation}
	P^{2M,s_{\rm \rm NL}}_{S,S_z,L,L_z,T,T_z}({\bf n}) =[a({\bf n}){\bf \nabla}^{2M}_{\rm 1/2}R^*_{\rm 
		L,L_{\rm z}}({\bf \nabla})a({\bf n})]^{s_{\rm \rm
			NL}}_{S,S_z,T,T_z}, 
\end{equation}
where ${\bf \nabla}_{\rm 1/2}^{2M}$ and ${\bf \nabla}$ act on the second annihilation operator.  This means we act on ${\bf n'}$ in Eq.~(\ref{spin-isospin}) and then set ${\bf n'}$ to equal
${\bf n}$.  The even integer $2M$ introduces extra derivatives. Writing the Clebsch-Gordan coefficients as $\langle S S_z, L L_z \vert J J_z\rangle$,
we define 
\begin{equation}
	O^{2M,s_{\rm NL}}_{S,L,J,J_z,T,T_z}({\bf n})= 
	\sum_{S_z,L_z}\langle S S_z, L L_z \vert J J_z\rangle P^{2M,s_{\rm NL}}_{S,S_z,L,L_z,T,T_z}({\bf
		n}).
\end{equation}
Using $O^{2M,s_{\rm NL}}_{S,L,J,J_z,T,T_z}({\bf n})$ and its Hermitian conjugate, $[O^{2M,s_{\rm NL}}_{S,L,J,J_z,T,T_z}({\bf n})]^\dagger$, we can construct short-range operators two-nucleon operators up to any order.  For the two-nucleon correlations operators used in this work, we simply 
take lowest order operators in all partial-wave channels.

\subsection{Correlations in the $T=1$ Channels}
\begin{figure}[h]
	\centering 
	\includegraphics[width=8cm]{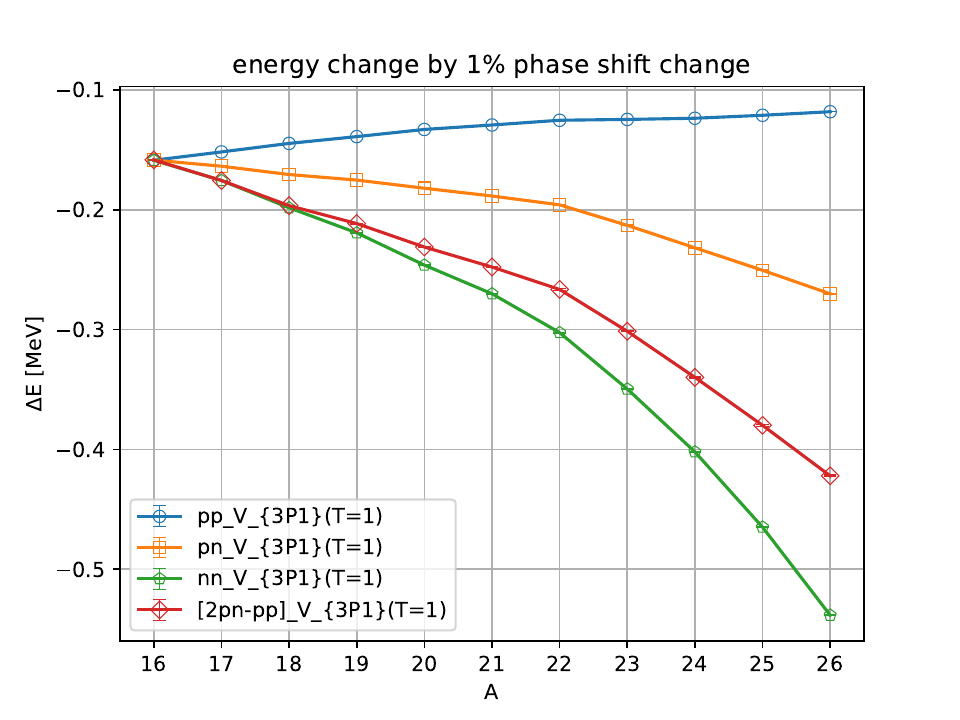}
	\includegraphics[width=8cm]{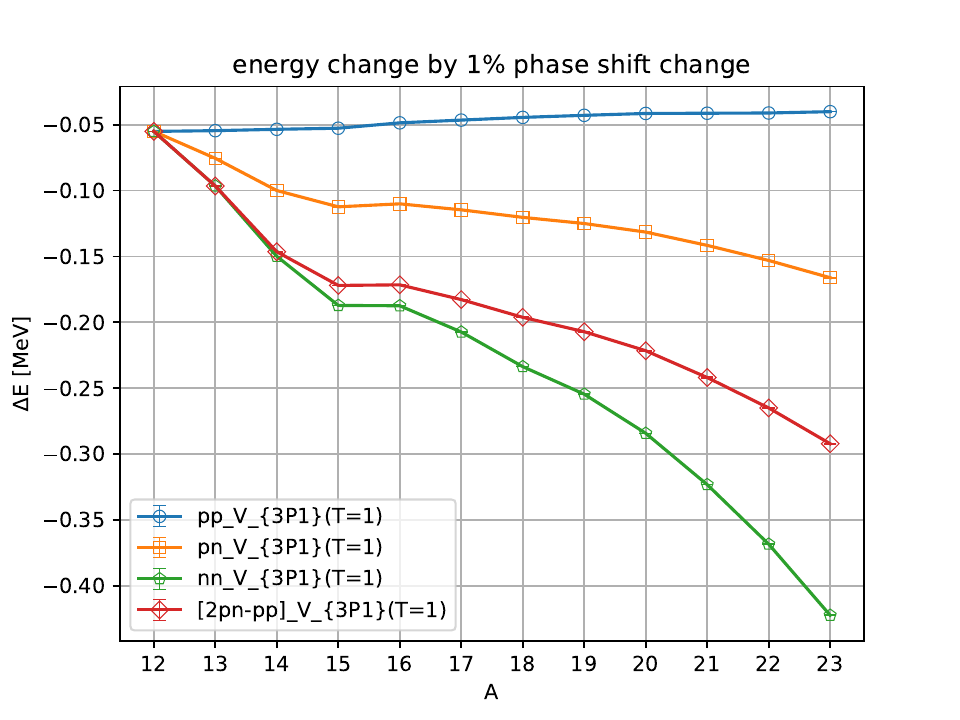}
	\caption{Correlations for pp, pn, nn, and 2pn$-$pp in the $^3$P$_1$ channel.  The left panel shows the oxygen isotopes, and the right panel shows the carbon isotopes.}
	\label{fig:3P1}
\end{figure}

\begin{figure}[h]
	\centering 
	\includegraphics[width=8cm]{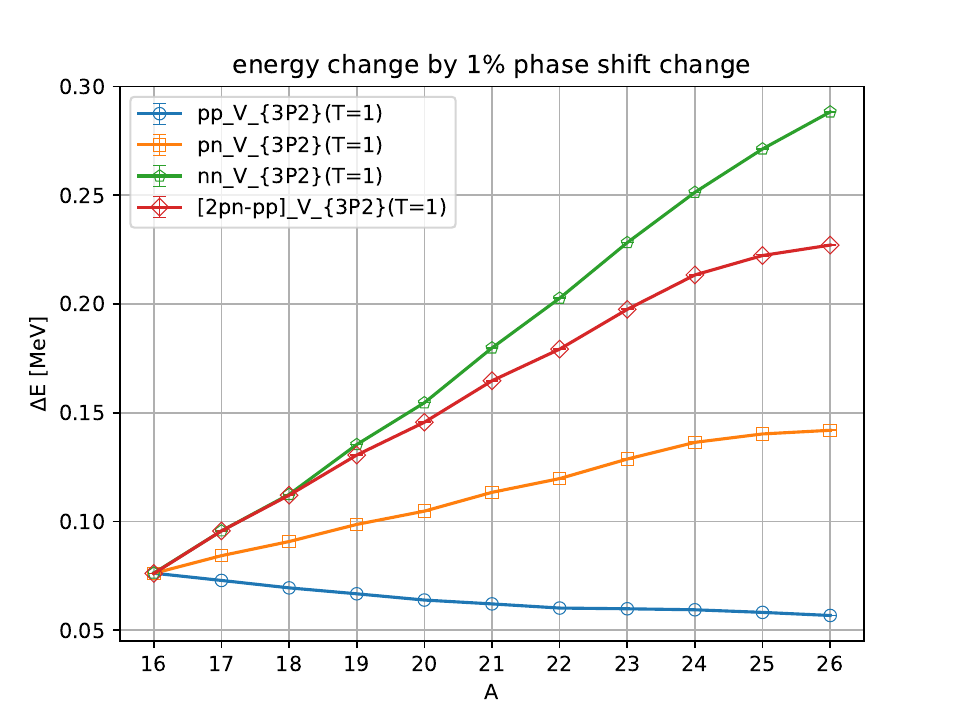}
	\includegraphics[width=8cm]{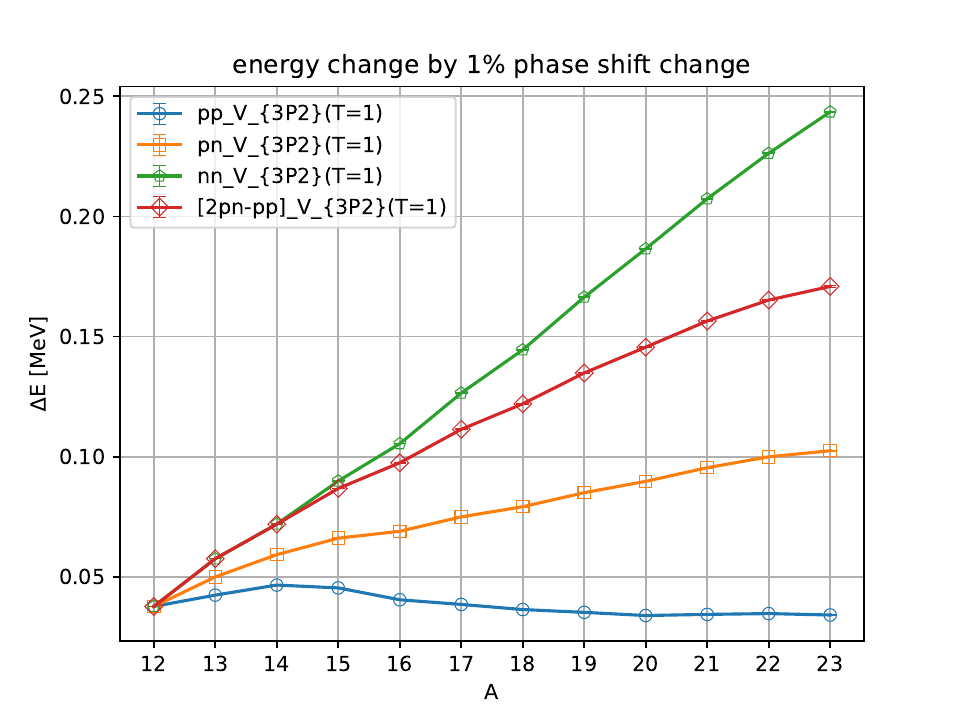}
	\caption{Correlations for pp, pn, nn, and 2pn$-$pp in the $^3$P$_2$ channel.  The left panel shows the oxygen isotopes, and the right panel shows the carbon isotopes.}
	\label{fig:3P2}
\end{figure}

\begin{figure}[h]
	\centering 
	\includegraphics[width=8cm]{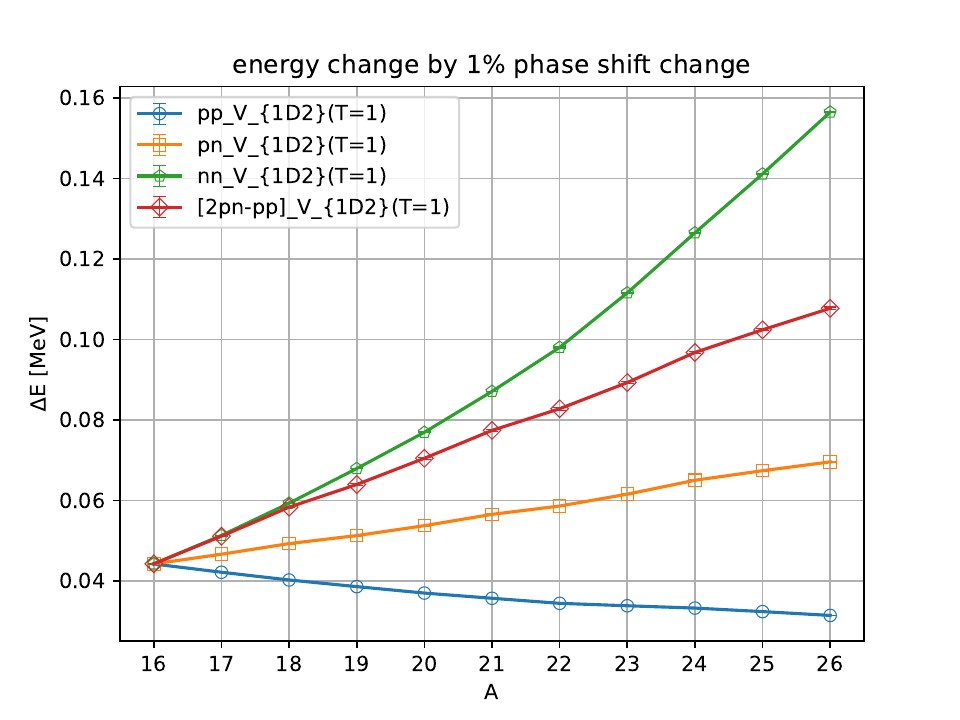}
	\includegraphics[width=8cm]{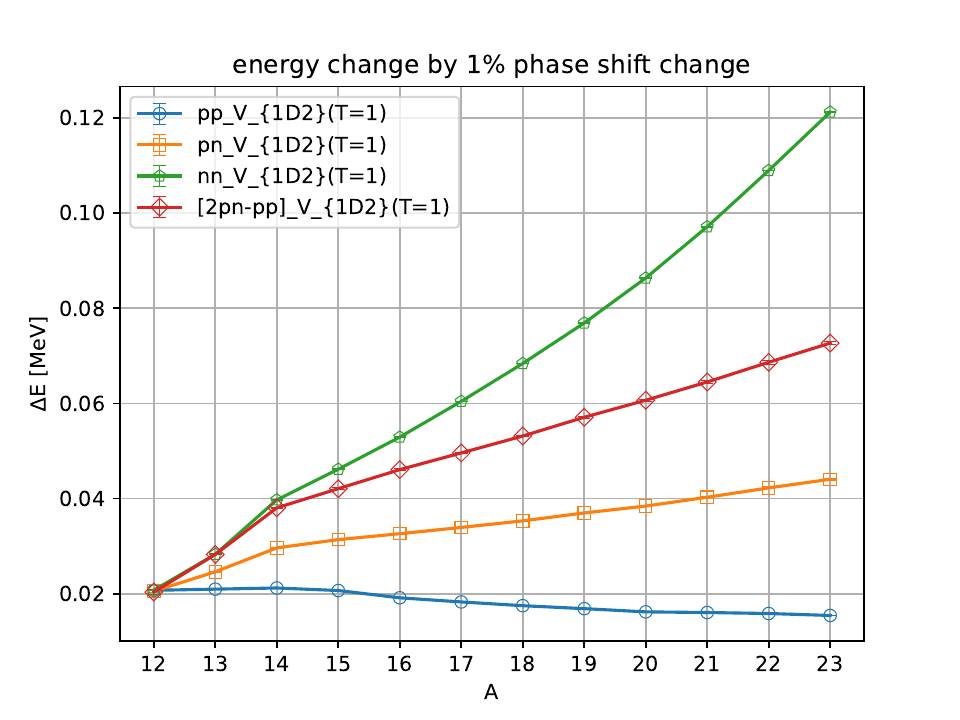}
	\caption{Correlations for pp, pn, nn, and 2pn$-$pp in the $^1$D$_2$ channel.  The left panel shows the oxygen isotopes, and the right panel shows the carbon isotopes.}
	\label{fig:1D2}
\end{figure}

In Fig.~\ref{fig:3P1}, we plot the correlations for pp, pn, nn, and 2pn$-$pp  in the $^3$P$_1$ channel, with the oxygen isotopes in the left panel and carbon isotopes in the right panel.  In Fig.~\ref{fig:3P2}, we plot the correlations for pp, pn, nn, and 2pn$-$pp  in the $^3$P$_2$ channel.  In Fig.~\ref{fig:1D2}, we plot the correlations for pp, pn, nn, and 2pn$-$pp  in the $^1$D$_2$ channel.

\subsection{Correlations in the $T=0$ Channels}

\begin{figure}[h]
	\centering 
	\includegraphics[width=8cm]{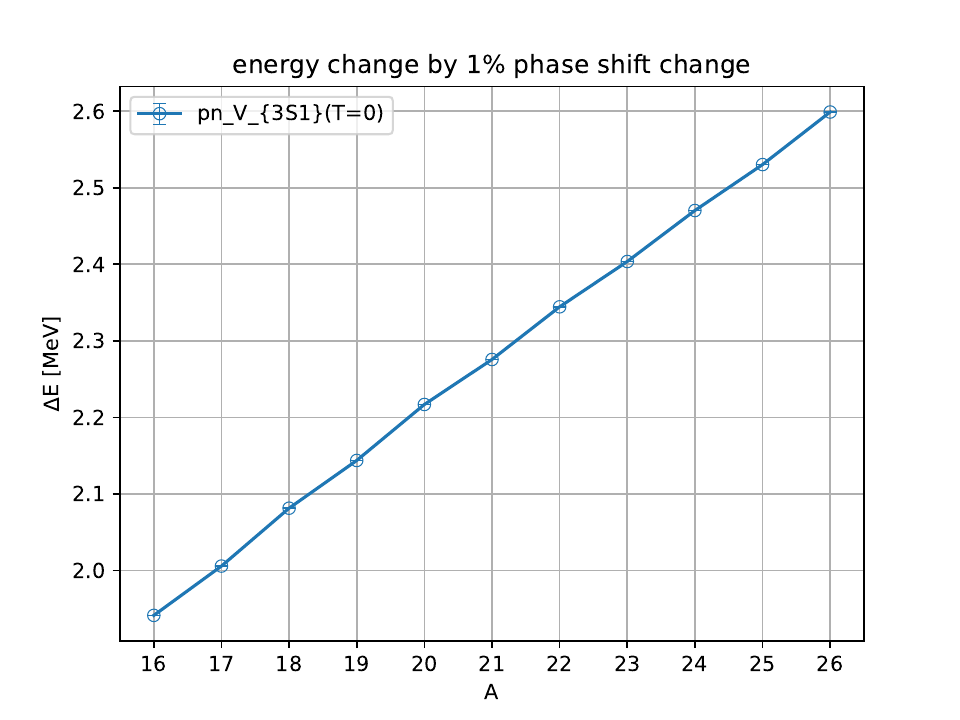}
	\includegraphics[width=8cm]
	{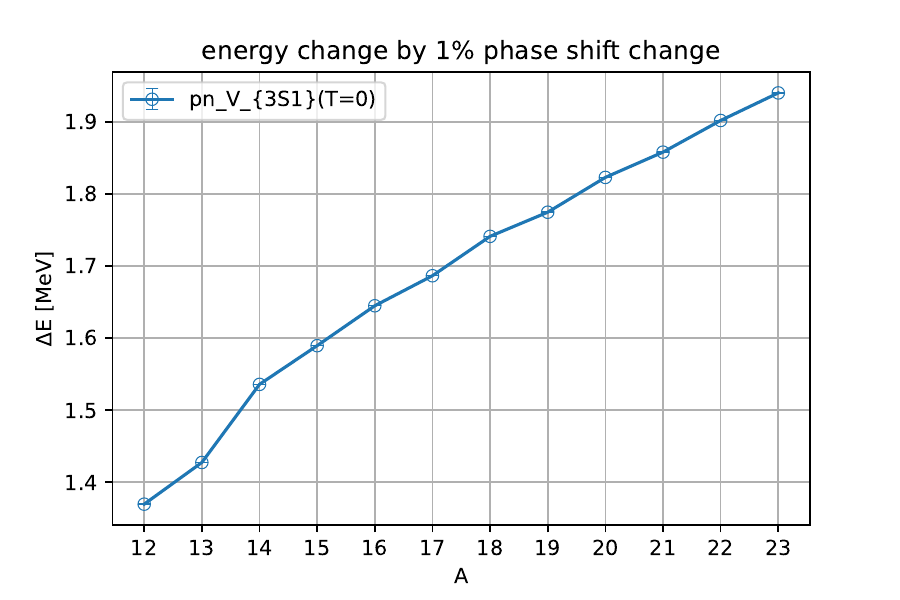}
	\caption{Correlations for pn in the $^3$S$_1$ channel.  The left panel shows the oxygen isotopes, and the right panel shows the carbon isotopes.}
	\label{fig:3S1}
\end{figure}

\begin{figure}[h]
	\centering 
	\includegraphics[width=8cm]{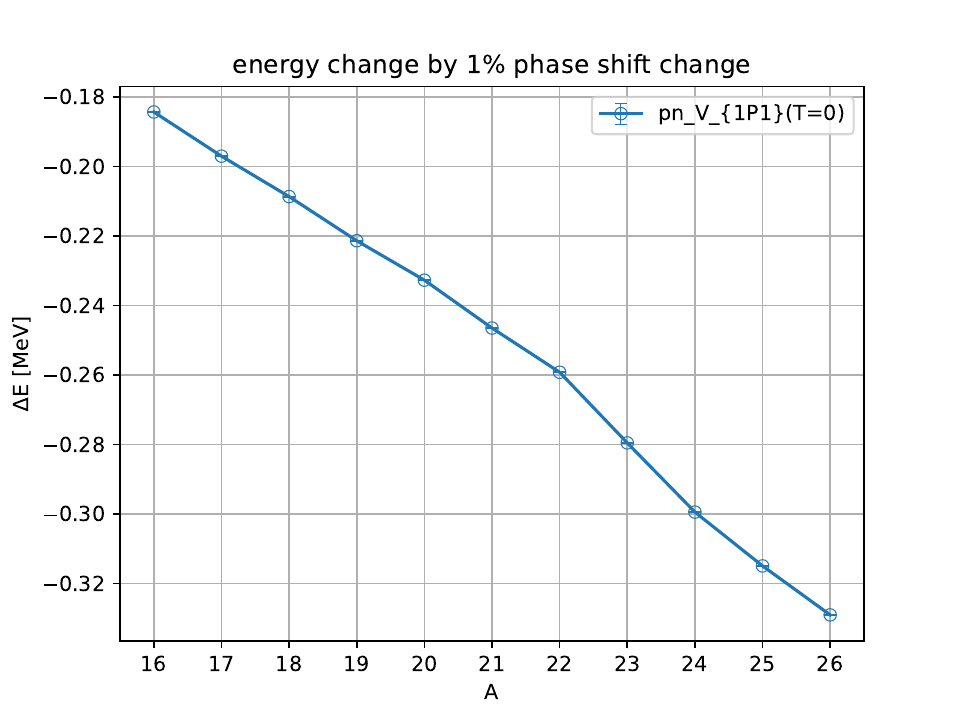}
	\includegraphics[width=8cm]
	{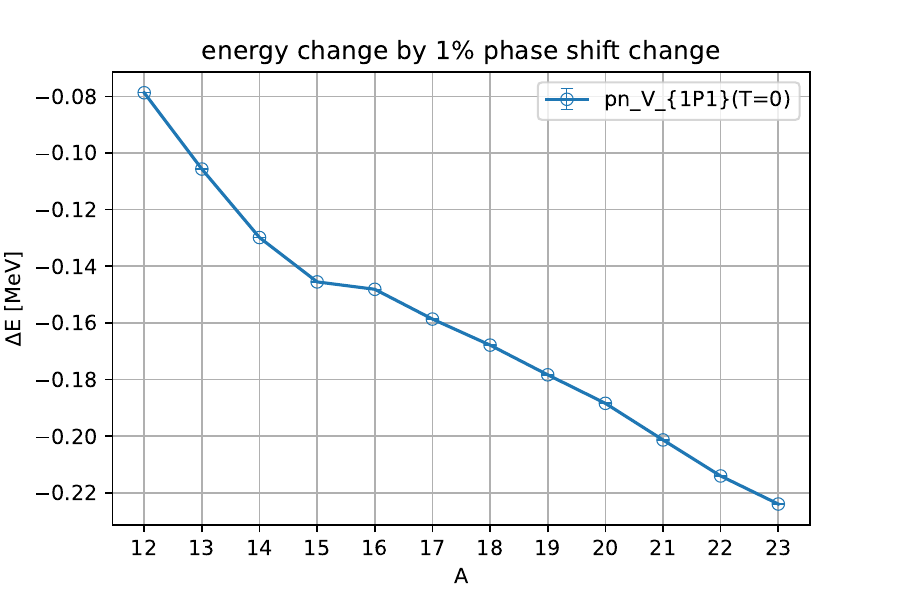}
	\caption{Correlations for pn in the $^1$P$_1$ channel.  The left panel shows the oxygen isotopes, and the right panel shows the carbon isotopes.}
	\label{fig:1P1}
\end{figure}

\begin{figure}[h]
	\centering 
	\includegraphics[width=8cm]{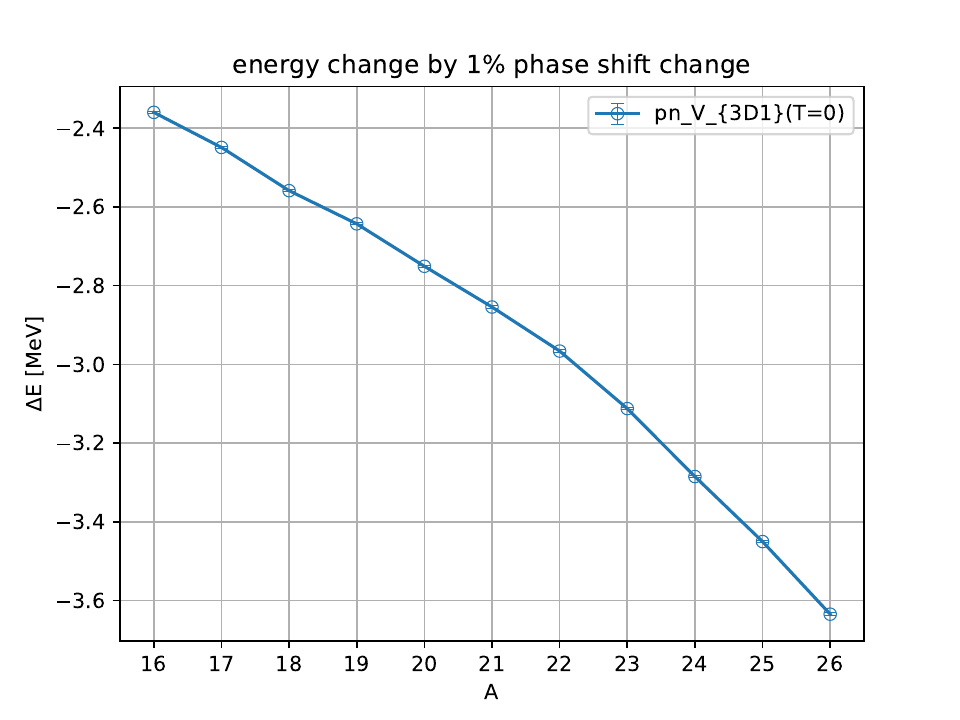}
	\includegraphics[width=8cm]
	{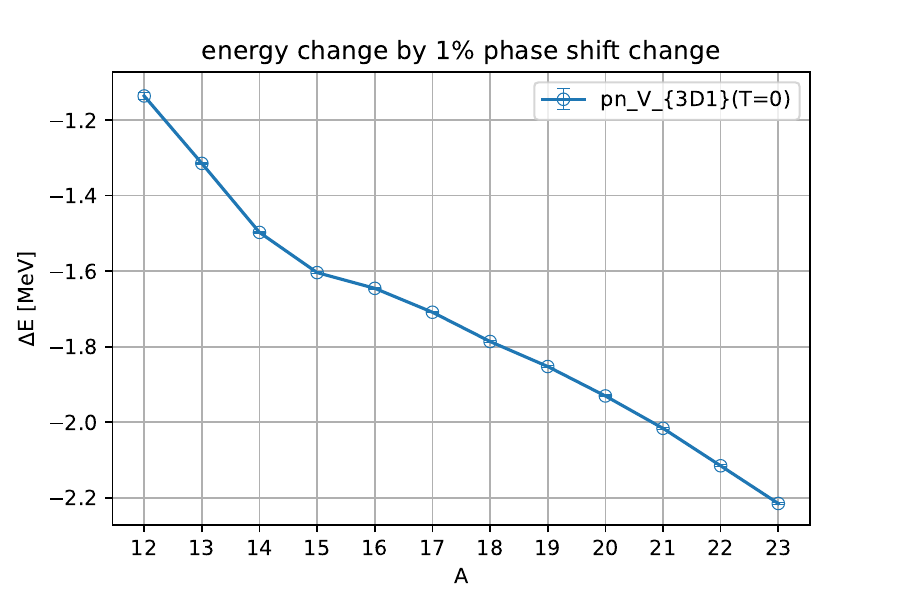}
	\caption{Correlations for pn in the $^3$D$_1$ channel.  The left panel shows the oxygen isotopes, and the right panel shows the carbon isotopes.}
	\label{fig:3D1}
\end{figure}

\begin{figure}[h]
	\centering 
	\includegraphics[width=8cm]{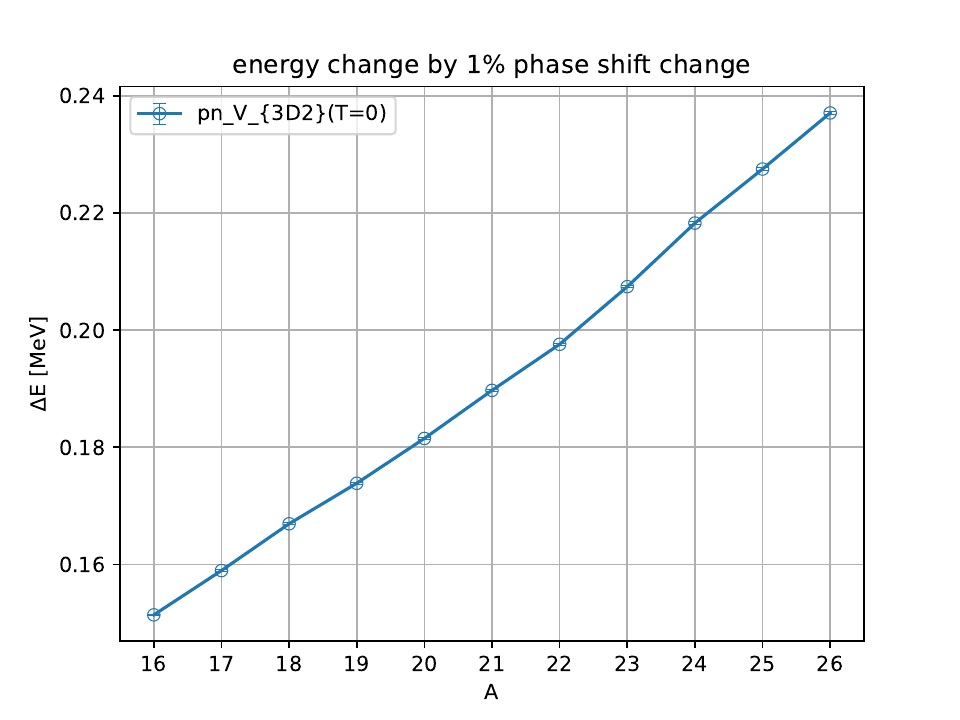}
	\includegraphics[width=8cm]
	{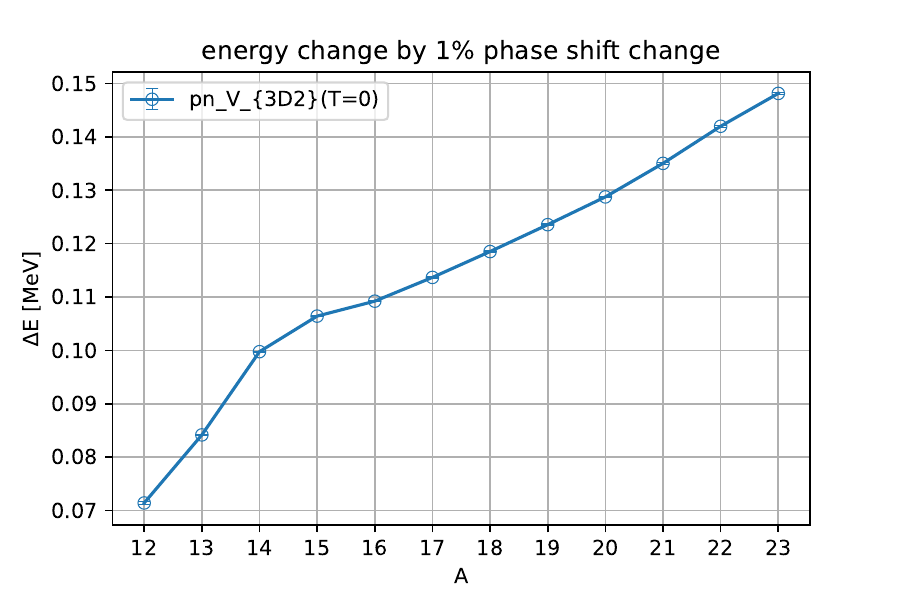}
	\caption{Correlations for pn in the $^3$D$_2$ channel.  The left panel shows the oxygen isotopes, and the right panel shows the carbon isotopes.}
	\label{fig:3D2}
\end{figure}

\begin{figure}[h]
	\centering 
	\includegraphics[width=8cm]{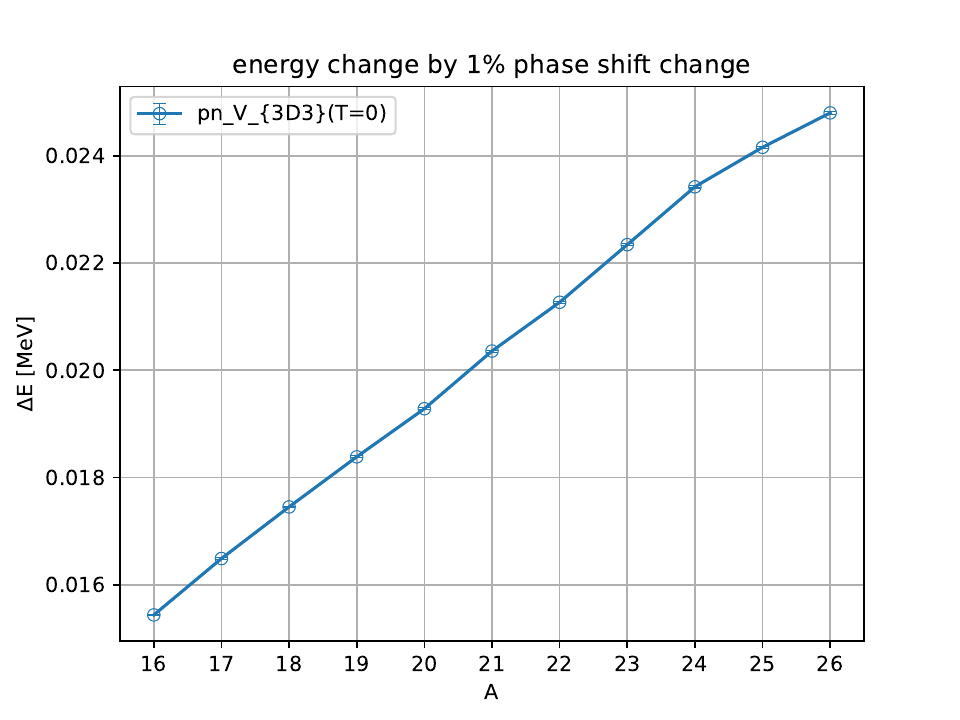}
	\includegraphics[width=8cm]
	{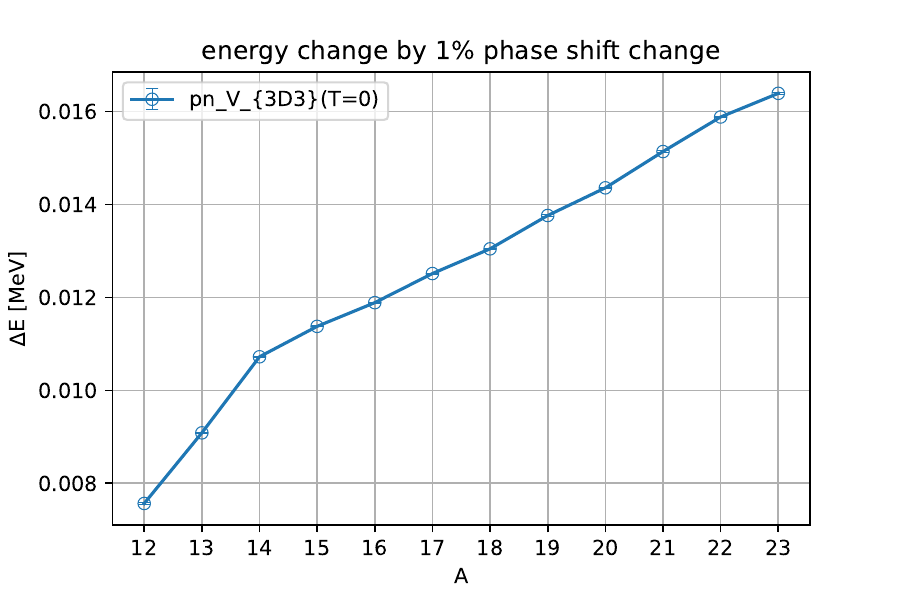}
	\caption{Correlations for pn in the $^3$D$_3$ channel.  The left panel shows the oxygen isotopes, and the right panel shows the carbon isotopes.}
	\label{fig:3D3}
\end{figure}

In Fig.~\ref{fig:3S1}, we plot the correlations for pn in the $^3$S$_1$ channel, with the oxygen isotopes in the left panel and carbon isotopes in the right panel.  In Fig.~\ref{fig:1P1}, we plot the correlations for pn in the $^1$P$_1$ channel.  In Fig.~\ref{fig:3D1}, we plot the correlations for pn in the $^3$D$_1$ channel.  In Fig.~\ref{fig:3D2}, we plot the correlations for pn in the $^3$D$_2$ channel. In Fig.~\ref{fig:3D3}, we plot the correlations for pn in the $^3$D$_3$ channel.

\subsection{Data for Energies}
The energies for the carbon isotopes are shown in Table~\ref{tab:C_energies} in comparison with experimental data.  The energies for the oxygen isotopes are shown in Table~\ref{tab:O_energies}.

\begin{table}[]
	\centering
	\begin{tabular}{|l|l|l|}
		\hline
		Nucleus & NLEFT (MeV) & Experiment (MeV) \\
		\hline
		\hline
		$^{12}$C ($0^+_1$) &  -92.4(6) & -92.16 \\
		$^{12}$C ($2^+_1$) &  -87.6(10) & -87.72 \\
		$^{12}$C ($0^+_2$) &  -84.9(14) & -84.51 \\
		$^{13}$C & -97.1(5) & -97.11 \\
		$^{14}$C & -104.8(7) & -105.28 \\
		$^{15}$C & -106.1(7) & -106.50 \\
		$^{16}$C & -111.1(7) & -110.75 \\
		$^{17}$C & -111.2(7) & -111.49 \\
		$^{18}$C & -116.3(7) & -115.67 \\
		$^{19}$C & -116.5(9) & -116.24 \\
		$^{20}$C & -120.0(13) & -119.22 \\
		$^{21}$C & -119.2(13) & -119.07 \\
		$^{22}$C & -120.4(13) & -119.26 \\
		$^{23}$C & -117.3(13) & -116.84 \\
		\hline
	\end{tabular}
	\caption{Energies for the carbon isotopes.
		Experimental data are from \cite{Kelley:2017qgh} \cite{Wang:2021xhn}.}
	\label{tab:C_energies}
\end{table}

\begin{table}[]
	\centering
	\begin{tabular}{|l|l|l|}
		\hline
		Nucleus & NLEFT (MeV) & Experiment (MeV) \\
		\hline
		\hline
		$^{16}$O & -130.0(4) & -127.62 \\
		$^{17}$O & -132.5(4) & -131.76 \\
		$^{18}$O & -140.4(5) & -139.81 \\
		$^{19}$O & -143.1(7) & -143.76 \\
		$^{20}$O & -151.9(13) & -151.37 \\
		$^{21}$O & -154.3(14) & -155.18 \\
		$^{22}$O & -160.6(17) & -162.03 \\
		$^{23}$O & -163.4(17) & -164.77 \\
		$^{24}$O & -166.9(17) & -168.38 \\
		$^{25}$O & -167.1(17) & -168.08 \\
		$^{26}$O & -169.4(17) & -167.88 \\
		\hline
	\end{tabular}
	\caption{Energies for the oxygen isotopes. 
		Experimental data are from \cite{Kelley:2017qgh} \cite{Wang:2021xhn}.}
	\label{tab:O_energies}
\end{table}

\subsection{Data for Phase Shifts}
For each partial wave, we show in Table~\ref{tab:phase_shifts} the phase shifts and the changes to the phase shifts produced by the two-nucleon operator perturbations.  We show the phase shifts at relative momenta $p=50$~MeV, $p=100$~MeV, and $p=150$~MeV.
\begin{table}
	\centering
	\begin{tabular}{|r|r|r|r|r|}
		\hline
		Channel & Momentum (MeV) & Phase Shift (deg) & New Phase Shift (deg) & Change \\
		\hline
		\hline
		$V_{\rm 1S0}$	& 50 & 6.356E+01 & 6.244E+01 & -1.76\% \\
		$V_{\rm 1S0}$	& 100 &	5.317E+01 &	5.257E+01 &	-1.13\% \\
		$V_{\rm 1S0}$	& 150 & 4.136E+01 &	4.094E+01 &	-1.00\% \\
		$V_{\rm 3P0}$	& 50 & 	1.823E+00 &	1.817E+00 &	-0.31\% \\
		$V_{\rm 3P0}$	& 100 & 7.039E+00 &	6.998E+00 &	-0.58\% \\
		$V_{\rm 3P0}$	& 150 & 1.003E+01 &	9.935E+00 &	-1.00\% \\
		$V_{\rm 3P1}$	& 50 & 	-1.120E+00 & -1.115E+00 & -0.44\% \\
		$V_{\rm 3P1}$	& 100 & -4.475E+00 & -4.443E+00	& -0.71\% \\
		$V_{\rm 3P1}$	& 150 & -8.091E+00 & -8.010E+00	& -1.01\% \\
		$V_{\rm 3P2}$	& 50 &	2.557E-01 &	2.537E-01 &	-0.80\% \\
		$V_{\rm 3P2}$	& 100 &	2.023E+00 &	2.005E+00 &	-0.85\% \\
		$V_{\rm 3P2}$	& 150 &	5.723E+00 &	5.666E+00 &	-1.00\% \\
		$V_{\rm 1D2}$	& 50 &	5.333E-02 &	5.327E-02 &	-0.12\% \\
		$V_{\rm 1D2}$	& 100 &	5.256E-01 &	5.236E-01 &	-0.37\% \\
		$V_{\rm 1D2}$	& 150 &	1.414E+00 &	1.400E+00 &	-1.00\% \\
		$V_{\rm 3S1}$	& 50 &	1.168E+02 &	1.161E+02 &	-0.64\% \\
		$V_{\rm 3S1}$	& 100 &	8.462E+01 &	8.393E+01 &	-0.82\% \\
		$V_{\rm 3S1}$	& 150 &	6.378E+01 &	6.314E+01 &	-1.00\% \\
		$V_{\rm 1P1}$	& 50 &	-1.683E+00 & -1.677E+00	& -0.37\% \\
		$V_{\rm 1P1}$	& 100 &	-5.905E+00 & -5.866E+00	& -0.67\% \\
		$V_{\rm 1P1}$	& 150 &	-9.738E+00 & -9.640E+00	& -1.01\% \\    
		$V_{\rm 3D1}$	& 50 &	-2.191E-01 & -2.169E-01	& -1.03\% \\
		$V_{\rm 3D1}$	& 100 &	-2.311E+00 & -2.280E+00	& -1.34\% \\
		$V_{\rm 3D1}$	& 150 &	-6.207E+00 & -6.145E+00	& -1.00\% \\
		$V_{\rm 3D2}$	& 50 &	2.596E-01 &	2.594E-01 &	-0.10\% \\
		$V_{\rm 3D2}$	& 100 &	2.859E+00 &	2.849E+00 &	-0.34\% \\
		$V_{\rm 3D2}$	& 150 &	7.581E+00 &	7.506E+00 &	-0.98\% \\
		$V_{\rm 3D3}$	& 50 &	6.410E-03 &	6.390E-03 &	-0.21\% \\
		$V_{\rm 3D3}$	& 100 &	3.245E-02 &	3.202E-02 &	-1.33\% \\
		$V_{\rm 3D3}$	& 150 &	3.070E-01 &	3.040E-01 &	-1.00\% \\
		\hline
	\end{tabular}
	\caption{Scattering phase shifts and phase shift changes produced by the two-nucleon operator perturbations}
	\label{tab:phase_shifts}
\end{table}

\subsection{Data for $T=1$ Correlations}
The data for the $^1$S$_0$ correlations are shown in Table~\ref{tab:1S0}.  The data for the $^3$P$_0$ correlations are in Table~\ref{tab:3P0}, $^3$P$_1$ correlations are in Table~\ref{tab:3P1}, $^3$P$_2$ correlations are in Table~\ref{tab:3P2}, and $^1$D$_2$ correlations are in Table~\ref{tab:1D2}.

\begin{table}
	\centering
	\begin{tabular}{|r|r|r|r|r|}
		\hline
		Nucleus	&	pp (MeV)	&	pn (MeV)	&	nn (MeV)	&	2pn$-$pp (MeV)	\\
		\hline
		\hline
		$^{16}$O	&	0.5060(1)	&	0.5061(1)	&	0.5061(1)	&	0.5061(2)	\\
		$^{17}$O	&	0.4963(1)	&	0.5227(1)	&	0.5496(1)	&	0.5492(2)	\\
		$^{18}$O	&	0.4866(1)	&	0.5427(2)	&	0.6465(1)	&	0.5988(3)	\\
		$^{19}$O	&	0.4781(1)	&	0.5587(2)	&	0.6929(1)	&	0.6394(3)	\\
		$^{20}$O	&	0.4695(1)	&	0.5779(2)	&	0.7925(2)	&	0.6863(3)	\\
		$^{21}$O	&	0.4626(1)	&	0.5934(2)	&	0.8385(1)	&	0.7243(3)	\\
		$^{22}$O	&	0.4559(1)	&	0.6117(3)	&	0.9388(2)	&	0.7675(5)	\\
		$^{23}$O	&	0.4520(1)	&	0.6269(3)	&	0.9847(2)	&	0.8017(5)	\\
		$^{24}$O	&	0.4481(1)	&	0.6440(3)	&	1.0776(2)	&	0.8398(4)	\\
		$^{25}$O	&	0.4429(1)	&	0.6598(3)	&	1.1307(2)	&	0.8767(4)	\\
		$^{26}$O	&	0.4376(1)	&	0.6772(3)	&	1.2317(2)	&	0.9168(5)	\\    
		\hline
	\end{tabular}
	\quad
	\quad
	\begin{tabular}{|r|r|r|r|r|}
		\hline
		Nucleus	&	pp (MeV)	&	pn (MeV)	&	nn (MeV)	&	2pn$-$pp (MeV)	\\
		\hline
		\hline
		$^{12}$C	&	0.3636(1)	&	0.3644(7)	&	0.3636(1)	&	0.3652(13)	\\
		$^{13}$C	&	0.3596(1)	&	0.3795(1)	&	0.3993(1)	&	0.3993(1)	\\
		$^{14}$C	&	0.3602(1)	&	0.4003(1)	&	0.5003(1)	&	0.4405(1)	\\
		$^{15}$C	&	0.3543(1)	&	0.4143(2)	&	0.5400(1)	&	0.4742(2)	\\
		$^{16}$C	&	0.3431(1)	&	0.4297(2)	&	0.6381(1)	&	0.5164(3)	\\
		$^{17}$C	&	0.3361(1)	&	0.4413(1)	&	0.6812(1)	&	0.5466(2)	\\
		$^{18}$C	&	0.3294(1)	&	0.4561(2)	&	0.7801(2)	&	0.5827(4)	\\
		$^{19}$C	&	0.3238(1)	&	0.4658(2)	&	0.8215(1)	&	0.6078(3)	\\
		$^{20}$C	&	0.3186(1)	&	0.4791(2)	&	0.9192(2)	&	0.6397(4)	\\
		$^{21}$C	&	0.3162(1)	&	0.4880(2)	&	0.9594(2)	&	0.6598(4)	\\
		$^{22}$C	&	0.3138(1)	&	0.4992(3)	&	1.0471(3)	&	0.6845(5)	\\
		$^{23}$C	&	0.3099(1)	&	0.5085(2)	&	1.0953(2)	&	0.7070(4)	\\    
		\hline
	\end{tabular}
	\caption{$^1$S$_0$ correlations for the oxygen and carbon isotopes}
	\label{tab:1S0}
\end{table}

\begin{table}
	\centering
	\begin{tabular}{|r|r|r|r|r|}
		\hline
		Nucleus	&	pp (MeV)	&	pn (MeV)	&	nn (MeV)	&	2pn$-$pp (MeV)	\\
		\hline
		\hline
		$^{16}$O	&	0.0353(1)	&	0.0353(1)	&	0.0353(1)	&	0.0353(1)	\\
		$^{17}$O	&	0.0338(1)	&	0.0351(1)	&	0.0365(1)	&	0.0365(1)	\\
		$^{18}$O	&	0.0322(1)	&	0.0359(1)	&	0.0395(1)	&	0.0397(1)	\\
		$^{19}$O	&	0.0309(1)	&	0.0357(1)	&	0.0408(1)	&	0.0404(1)	\\
		$^{20}$O	&	0.0296(1)	&	0.0365(1)	&	0.0445(1)	&	0.0435(1)	\\
		$^{21}$O	&	0.0288(1)	&	0.0367(1)	&	0.0458(1)	&	0.0446(1)	\\
		$^{22}$O	&	0.0279(1)	&	0.0377(1)	&	0.0503(1)	&	0.0475(1)	\\
		$^{23}$O	&	0.0277(1)	&	0.0413(1)	&	0.0602(1)	&	0.0549(1)	\\
		$^{24}$O	&	0.0275(1)	&	0.0458(1)	&	0.0726(1)	&	0.0641(1)	\\
		$^{25}$O	&	0.0270(1)	&	0.0512(1)	&	0.0905(1)	&	0.0754(1)	\\
		$^{26}$O	&	0.0263(1)	&	0.0574(1)	&	0.1119(1)	&	0.0884(1)	\\   
		\hline
	\end{tabular}
	\quad
	\quad
	\begin{tabular}{|r|r|r|r|r|}
		\hline
		Nucleus	&	pp (MeV)	&	pn (MeV)	&	nn (MeV)	&	2pn$-$pp (MeV)	\\
		\hline
		\hline
		$^{12}$C	&	0.0093(1)	&	0.0093(1)	&	0.0093(1)	&	0.0093(1)	\\
		$^{13}$C	&	0.0077(1)	&	0.0133(1)	&	0.0189(1)	&	0.0189(1)	\\
		$^{14}$C	&	0.0059(1)	&	0.0197(1)	&	0.0334(1)	&	0.0334(1)	\\
		$^{15}$C	&	0.0060(1)	&	0.0222(1)	&	0.0417(1)	&	0.0384(1)	\\
		$^{16}$C	&	0.0060(1)	&	0.0203(1)	&	0.0376(1)	&	0.0347(1)	\\
		$^{17}$C	&	0.0058(1)	&	0.0202(1)	&	0.0391(1)	&	0.0346(1)	\\
		$^{18}$C	&	0.0057(1)	&	0.0211(1)	&	0.0431(1)	&	0.0364(1)	\\
		$^{19}$C	&	0.0054(1)	&	0.0209(1)	&	0.0443(1)	&	0.0364(1)	\\
		$^{20}$C	&	0.0053(1)	&	0.0219(1)	&	0.0488(1)	&	0.0385(1)	\\
		$^{21}$C	&	0.0051(1)	&	0.0239(1)	&	0.0571(1)	&	0.0428(1)	\\
		$^{22}$C	&	0.0049(1)	&	0.0269(1)	&	0.0679(1)	&	0.0488(1)	\\
		$^{23}$C	&	0.0047(1)	&	0.0308(1)	&	0.0832(1)	&	0.0568(1)	\\   
		\hline
	\end{tabular}
	\caption{$^3$P$_0$ correlations for the oxygen and carbon isotopes}
	\label{tab:3P0}
\end{table}

\begin{table}
	\centering
	\begin{tabular}{|r|r|r|r|r|}
		\hline
		Nucleus	&	pp (MeV)	&	pn (MeV)	&	nn (MeV)	&	2pn$-$pp (MeV)	\\
		\hline
		\hline
		$^{16}$O	&	-0.1587(1)	&	-0.1585(2)	&	-0.1587(1)	&	-0.1584(2)	\\
		$^{17}$O	&	-0.1517(1)	&	-0.1636(2)	&	-0.1756(1)	&	-0.1756(3)	\\
		$^{18}$O	&	-0.1446(1)	&	-0.1706(2)	&	-0.1983(1)	&	-0.1966(3)	\\
		$^{19}$O	&	-0.1389(1)	&	-0.1752(2)	&	-0.2193(1)	&	-0.2115(3)	\\
		$^{20}$O	&	-0.1329(1)	&	-0.1820(2)	&	-0.2462(2)	&	-0.2311(4)	\\
		$^{21}$O	&	-0.1292(1)	&	-0.1885(3)	&	-0.2701(2)	&	-0.2478(4)	\\
		$^{22}$O	&	-0.1253(1)	&	-0.1959(3)	&	-0.3026(3)	&	-0.2665(6)	\\
		$^{23}$O	&	-0.1246(1)	&	-0.2130(3)	&	-0.3496(2)	&	-0.3014(5)	\\
		$^{24}$O	&	-0.1236(1)	&	-0.2318(3)	&	-0.4021(2)	&	-0.3399(5)	\\
		$^{25}$O	&	-0.1211(1)	&	-0.2505(4)	&	-0.4650(3)	&	-0.3800(6)	\\
		$^{26}$O	&	-0.1182(1)	&	-0.2701(4)	&	-0.5385(3)	&	-0.4221(7)	\\
		\hline
	\end{tabular}
	\quad
	\quad
	\begin{tabular}{|r|r|r|r|r|}
		\hline
		Nucleus	&	pp (MeV)	&	pn (MeV)	&	nn (MeV)	&	2pn$-$pp (MeV)	\\
		\hline
		\hline
		$^{12}$C	&	-0.0551(1)	&	-0.0551(1)	&	-0.0551(1)	&	-0.0551(1)	\\
		$^{13}$C	&	-0.0545(1)	&	-0.0755(1)	&	-0.0965(1)	&	-0.0965(1)	\\
		$^{14}$C	&	-0.0535(1)	&	-0.1000(1)	&	-0.1499(1)	&	-0.1465(2)	\\
		$^{15}$C	&	-0.0527(1)	&	-0.1123(2)	&	-0.1872(1)	&	-0.1720(2)	\\
		$^{16}$C	&	-0.0486(1)	&	-0.1101(1)	&	-0.1875(1)	&	-0.1716(2)	\\
		$^{17}$C	&	-0.0465(1)	&	-0.1146(1)	&	-0.2074(1)	&	-0.1828(2)	\\
		$^{18}$C	&	-0.0445(1)	&	-0.1203(2)	&	-0.2337(2)	&	-0.1962(4)	\\
		$^{19}$C	&	-0.0429(1)	&	-0.1250(3)	&	-0.2546(2)	&	-0.2072(5)	\\
		$^{20}$C	&	-0.0415(1)	&	-0.1315(2)	&	-0.2843(2)	&	-0.2216(3)	\\
		$^{21}$C	&	-0.0413(1)	&	-0.1416(2)	&	-0.3233(2)	&	-0.2419(4)	\\
		$^{22}$C	&	-0.0411(1)	&	-0.1531(3)	&	-0.3684(2)	&	-0.2650(5)	\\
		$^{23}$C	&	-0.0401(1)	&	-0.1662(3)	&	-0.4224(2)	&	-0.2922(5)	\\
		\hline
	\end{tabular}
	\caption{$^3$P$_1$ correlations for the oxygen and carbon isotopes}
	\label{tab:3P1}
\end{table}

\begin{table}
	\centering
	\begin{tabular}{|r|r|r|r|r|}
		\hline
		Nucleus	&	pp (MeV)	&	pn (MeV)	&	nn (MeV)	&	2pn$-$pp (MeV)	\\
		\hline
		\hline
		$^{16}$O	&	0.0762(1)	&	0.0761(1)	&	0.0762(1)	&	0.0761(1)	\\
		$^{17}$O	&	0.0729(1)	&	0.0842(1)	&	0.0956(1)	&	0.0956(1)	\\
		$^{18}$O	&	0.0694(1)	&	0.0907(1)	&	0.1124(1)	&	0.1120(2)	\\
		$^{19}$O	&	0.0667(1)	&	0.0986(1)	&	0.1353(1)	&	0.1305(2)	\\
		$^{20}$O	&	0.0638(1)	&	0.1047(1)	&	0.1546(1)	&	0.1456(2)	\\
		$^{21}$O	&	0.0621(1)	&	0.1134(1)	&	0.1798(1)	&	0.1647(2)	\\
		$^{22}$O	&	0.0602(1)	&	0.1197(2)	&	0.2026(2)	&	0.1792(4)	\\
		$^{23}$O	&	0.0598(1)	&	0.1287(2)	&	0.2282(1)	&	0.1975(3)	\\
		$^{24}$O	&	0.0594(1)	&	0.1364(2)	&	0.2513(1)	&	0.2133(3)	\\
		$^{25}$O	&	0.0582(1)	&	0.1402(2)	&	0.2713(2)	&	0.2223(3)	\\
		$^{26}$O	&	0.0567(1)	&	0.1419(2)	&	0.2883(2)	&	0.2271(3)	\\
		\hline
	\end{tabular}
	\quad
	\quad
	\begin{tabular}{|r|r|r|r|r|}
		\hline
		Nucleus	&	pp (MeV)	&	pn (MeV)	&	nn (MeV)	&	2pn$-$pp (MeV)	\\
		\hline
		\hline
		$^{12}$C	&	0.0377(1)	&	0.0377(1)	&	0.0377(1)	&	0.0377(1)	\\
		$^{13}$C	&	0.0424(1)	&	0.0500(1)	&	0.0576(1)	&	0.0576(1)	\\
		$^{14}$C	&	0.0466(1)	&	0.0593(1)	&	0.0720(1)	&	0.0719(1)	\\
		$^{15}$C	&	0.0454(1)	&	0.0661(1)	&	0.0899(1)	&	0.0868(1)	\\
		$^{16}$C	&	0.0405(1)	&	0.0690(1)	&	0.1054(1)	&	0.0975(1)	\\
		$^{17}$C	&	0.0386(1)	&	0.0750(1)	&	0.1265(1)	&	0.1114(1)	\\
		$^{18}$C	&	0.0364(1)	&	0.0792(2)	&	0.1445(1)	&	0.1220(3)	\\
		$^{19}$C	&	0.0352(1)	&	0.0850(1)	&	0.1664(1)	&	0.1349(2)	\\
		$^{20}$C	&	0.0339(1)	&	0.0898(1)	&	0.1865(1)	&	0.1456(2)	\\
		$^{21}$C	&	0.0344(1)	&	0.0954(1)	&	0.2073(1)	&	0.1565(2)	\\
		$^{22}$C	&	0.0347(1)	&	0.1000(1)	&	0.2263(1)	&	0.1652(3)	\\
		$^{23}$C	&	0.0341(1)	&	0.1025(1)	&	0.2435(1)	&	0.1708(3)	\\   
		\hline
	\end{tabular}
	\caption{$^3$P$_2$ correlations for the oxygen and carbon isotopes}
	\label{tab:3P2}
\end{table}

\begin{table}
	\centering
	\begin{tabular}{|r|r|r|r|r|}
		\hline
		Nucleus	&	pp (MeV)	&	pn (MeV)	&	nn (MeV)	&	2pn$-$pp (MeV)	\\
		\hline
		\hline
		$^{16}$O	&	0.0442(1)	&	0.0442(1)	&	0.0442(1)	&	0.0442(1)	\\
		$^{17}$O	&	0.0421(1)	&	0.0466(1)	&	0.0513(1)	&	0.0511(2)	\\
		$^{18}$O	&	0.0402(1)	&	0.0493(1)	&	0.0593(1)	&	0.0583(2)	\\
		$^{19}$O	&	0.0386(1)	&	0.0513(1)	&	0.0679(1)	&	0.0639(2)	\\
		$^{20}$O	&	0.0370(1)	&	0.0537(1)	&	0.0769(1)	&	0.0705(2)	\\
		$^{21}$O	&	0.0357(1)	&	0.0565(1)	&	0.0871(1)	&	0.0774(2)	\\
		$^{22}$O	&	0.0344(1)	&	0.0586(2)	&	0.0980(1)	&	0.0828(3)	\\
		$^{23}$O	&	0.0338(1)	&	0.0615(2)	&	0.1115(1)	&	0.0893(3)	\\
		$^{24}$O	&	0.0332(1)	&	0.0650(2)	&	0.1265(1)	&	0.0967(3)	\\
		$^{25}$O	&	0.0324(1)	&	0.0674(2)	&	0.1411(1)	&	0.1024(3)	\\
		$^{26}$O	&	0.0314(1)	&	0.0696(2)	&	0.1564(1)	&	0.1077(3)	\\   
		\hline
	\end{tabular}
	\quad
	\quad
	\begin{tabular}{|r|r|r|r|r|}
		\hline
		Nucleus	&	pp (MeV)	&	pn (MeV)	&	nn (MeV)	&	2pn$-$pp (MeV)	\\
		\hline
		\hline
		$^{12}$C	&	0.0207(1)	&	0.0206(3)	&	0.0208(1)	&	0.0204(5)	\\
		$^{13}$C	&	0.0210(1)	&	0.0246(1)	&	0.0283(1)	&	0.0283(1)	\\
		$^{14}$C	&	0.0212(1)	&	0.0296(1)	&	0.0397(1)	&	0.0381(1)	\\
		$^{15}$C	&	0.0207(1)	&	0.0314(1)	&	0.0462(1)	&	0.0421(1)	\\
		$^{16}$C	&	0.0192(1)	&	0.0326(1)	&	0.0529(1)	&	0.0461(1)	\\
		$^{17}$C	&	0.0183(1)	&	0.0339(1)	&	0.0604(1)	&	0.0496(1)	\\
		$^{18}$C	&	0.0175(1)	&	0.0353(1)	&	0.0684(1)	&	0.0531(2)	\\
		$^{19}$C	&	0.0169(1)	&	0.0370(1)	&	0.0769(1)	&	0.0571(2)	\\
		$^{20}$C	&	0.0162(1)	&	0.0384(1)	&	0.0863(1)	&	0.0607(3)	\\
		$^{21}$C	&	0.0161(1)	&	0.0403(1)	&	0.0971(1)	&	0.0645(2)	\\
		$^{22}$C	&	0.0159(1)	&	0.0423(2)	&	0.1089(1)	&	0.0686(4)	\\
		$^{23}$C	&	0.0155(1)	&	0.0441(1)	&	0.1212(1)	&	0.0727(3)	\\   
		\hline
	\end{tabular}
	\caption{$^1$D$_2$ correlations for the oxygen and carbon isotopes}
	\label{tab:1D2}
\end{table}

\subsection{Data for $T=0$ Correlations}
The data for the $^3$S$_1$ correlations are shown in Table~\ref{tab:3S1}.  The data for the $^1$P$_1$ correlations are in Table~\ref{tab:1P1}, $^3$D$_1$ correlations are in Table~\ref{tab:3D1}, $^3$D$_2$ correlations are in Table~\ref{tab:3D2}, and $^3$D$_3$ correlations are in Table~\ref{tab:3D3}.

\begin{table}
	\centering
	\begin{tabular}{|r|r|}
		\hline
		Nucleus	&	pn (MeV)	\\
		\hline
		\hline
		$^{16}$O	&	1.9414(3)	\\
		$^{17}$O	&	2.0058(3)	\\
		$^{18}$O	&	2.0814(4)	\\
		$^{19}$O	&	2.1437(4)	\\
		$^{20}$O	&	2.2169(4)	\\
		$^{21}$O	&	2.2757(4)	\\
		$^{22}$O	&	2.3444(5)	\\
		$^{23}$O	&	2.4037(4)	\\
		$^{24}$O	&	2.4703(5)	\\
		$^{25}$O	&	2.5303(5)	\\
		$^{26}$O	&	2.5989(5)	\\
		\hline
	\end{tabular}
	\quad
	\quad
	\begin{tabular}{|r|r|}
		\hline
		Nucleus	&	pn (MeV)	\\
		\hline
		\hline
		$^{12}$C	&	1.3696(4)	\\
		$^{13}$C	&	1.4274(2)	\\
		$^{14}$C	&	1.5357(2)	\\
		$^{15}$C	&	1.5894(3)	\\
		$^{16}$C	&	1.6448(3)	\\
		$^{17}$C	&	1.6864(3)	\\
		$^{18}$C	&	1.7410(3)	\\
		$^{19}$C	&	1.7745(3)	\\
		$^{20}$C	&	1.8228(3)	\\
		$^{21}$C	&	1.8578(4)	\\
		$^{22}$C	&	1.9017(4)	\\
		$^{23}$C	&	1.9400(4)	\\
		\hline
	\end{tabular}
	\caption{$^3$S$_1$ correlations for the oxygen and carbon isotopes}
	\label{tab:3S1}
\end{table}

\begin{table}
	\centering
	\begin{tabular}{|r|r|}
		\hline
		Nucleus	&	pn (MeV)	\\
		\hline
		\hline
		$^{16}$O	&	-0.1843(1)	\\
		$^{17}$O	&	-0.1970(1)	\\
		$^{18}$O	&	-0.2087(1)	\\
		$^{19}$O	&	-0.2214(1)	\\
		$^{20}$O	&	-0.2327(1)	\\
		$^{21}$O	&	-0.2465(2)	\\
		$^{22}$O	&	-0.2592(2)	\\
		$^{23}$O	&	-0.2795(2)	\\
		$^{24}$O	&	-0.2994(2)	\\
		$^{25}$O	&	-0.3149(2)	\\
		$^{26}$O	&	-0.3290(2)	\\
		\hline
	\end{tabular}
	\quad
	\quad
	\begin{tabular}{|r|r|}
		\hline
		Nucleus	&	pn (MeV)	\\
		\hline
		\hline
		$^{12}$C	&	-0.0787(1)	\\
		$^{13}$C	&	-0.1057(1)	\\
		$^{14}$C	&	-0.1299(1)	\\
		$^{15}$C	&	-0.1455(1)	\\
		$^{16}$C	&	-0.1481(1)	\\
		$^{17}$C	&	-0.1586(1)	\\
		$^{18}$C	&	-0.1678(1)	\\
		$^{19}$C	&	-0.1783(1)	\\
		$^{20}$C	&	-0.1884(1)	\\
		$^{21}$C	&	-0.2013(1)	\\
		$^{22}$C	&	-0.2140(1)	\\
		$^{23}$C	&	-0.2239(1)	\\
		\hline
	\end{tabular}
	\caption{$^1$P$_1$ correlations for the oxygen and carbon isotopes}
	\label{tab:1P1}
\end{table}

\begin{table}
	\centering
	\begin{tabular}{|r|r|}
		\hline
		Nucleus	&	pn (MeV)	\\
		\hline
		\hline
		$^{16}$O	&	-2.3597(19)	\\
		$^{17}$O	&	-2.4487(23)	\\
		$^{18}$O	&	-2.5586(25)	\\
		$^{19}$O	&	-2.6429(26)	\\
		$^{20}$O	&	-2.7508(27)	\\
		$^{21}$O	&	-2.8543(29)	\\
		$^{22}$O	&	-2.9662(30)	\\
		$^{23}$O	&	-3.1124(31)	\\
		$^{24}$O	&	-3.2851(36)	\\
		$^{25}$O	&	-3.4504(33)	\\
		$^{26}$O	&	-3.6348(36)	\\   
		\hline
	\end{tabular}
	\quad
	\quad
	\begin{tabular}{|r|r|}
		\hline
		Nucleus	&	pn (MeV)	\\
		\hline
		\hline
		$^{12}$C	&	-1.1359(90)	\\
		$^{13}$C	&	-1.3145(16)	\\
		$^{14}$C	&	-1.4974(15)	\\
		$^{15}$C	&	-1.6038(19)	\\
		$^{16}$C	&	-1.6455(18)	\\
		$^{17}$C	&	-1.7088(18)	\\
		$^{18}$C	&	-1.7863(18)	\\
		$^{19}$C	&	-1.8526(22)	\\
		$^{20}$C	&	-1.9305(25)	\\
		$^{21}$C	&	-2.0163(23)	\\
		$^{22}$C	&	-2.1155(25)	\\
		$^{23}$C	&	-2.2153(25)	\\    
		\hline
	\end{tabular}
	\caption{$^3$D$_1$ correlations for the oxygen and carbon isotopes}
	\label{tab:3D1}
\end{table}    

\begin{table}
	\centering
	\begin{tabular}{|r|r|}
		\hline
		Nucleus	&	pn (MeV)	\\
		\hline
		\hline
		$^{16}$O	&	0.1514(1)	\\
		$^{17}$O	&	0.1589(1)	\\
		$^{18}$O	&	0.1669(2)	\\
		$^{19}$O	&	0.1739(2)	\\
		$^{20}$O	&	0.1815(2)	\\
		$^{21}$O	&	0.1897(2)	\\
		$^{22}$O	&	0.1976(2)	\\
		$^{23}$O	&	0.2074(2)	\\
		$^{24}$O	&	0.2183(2)	\\
		$^{25}$O	&	0.2275(2)	\\
		$^{26}$O	&	0.2371(2)	\\   
		\hline
	\end{tabular}
	\quad
	\quad
	\begin{tabular}{|r|r|}
		\hline
		Nucleus	&	pn (MeV)	\\
		\hline
		\hline
		$^{12}$C	&	0.0714(3)	\\
		$^{13}$C	&	0.0841(1)	\\
		$^{14}$C	&	0.0998(1)	\\
		$^{15}$C	&	0.1064(1)	\\
		$^{16}$C	&	0.1092(1)	\\
		$^{17}$C	&	0.1137(1)	\\
		$^{18}$C	&	0.1185(1)	\\
		$^{19}$C	&	0.1236(1)	\\
		$^{20}$C	&	0.1288(1)	\\
		$^{21}$C	&	0.1351(1)	\\
		$^{22}$C	&	0.1420(1)	\\
		$^{23}$C	&	0.1482(2)	\\
		\hline
	\end{tabular}
	\caption{$^3$D$_2$ correlations for the oxygen and carbon isotopes}
	\label{tab:3D2}
\end{table} 

\begin{table}
	\centering
	\begin{tabular}{|r|r|}
		\hline
		Nucleus	&	pn (MeV)	\\
		\hline
		\hline
		$^{16}$O	&	0.0154(1)	\\
		$^{17}$O	&	0.0165(1)	\\
		$^{18}$O	&	0.0175(1)	\\
		$^{19}$O	&	0.0184(1)	\\
		$^{20}$O	&	0.0193(1)	\\
		$^{21}$O	&	0.0204(1)	\\
		$^{22}$O	&	0.0213(1)	\\
		$^{23}$O	&	0.0223(1)	\\
		$^{24}$O	&	0.0234(1)	\\
		$^{25}$O	&	0.0242(1)	\\
		$^{26}$O	&	0.0248(1)	\\
		\hline
	\end{tabular}
	\quad
	\quad
	\begin{tabular}{|r|r|}
		\hline
		Nucleus	&	pn (MeV)	\\
		\hline
		\hline
		$^{12}$C	&	0.0076(1)	\\
		$^{13}$C	&	0.0091(1)	\\
		$^{14}$C	&	0.0107(1)	\\
		$^{15}$C	&	0.0114(1)	\\
		$^{16}$C	&	0.0119(1)	\\
		$^{17}$C	&	0.0125(1)	\\
		$^{18}$C	&	0.0130(1)	\\
		$^{19}$C	&	0.0138(1)	\\
		$^{20}$C	&	0.0144(1)	\\
		$^{21}$C	&	0.0151(1)	\\
		$^{22}$C	&	0.0159(1)	\\
		$^{23}$C	&	0.0164(1)	\\
		\hline
	\end{tabular}
	\caption{$^3$D$_3$ correlations for the oxygen and carbon isotopes}
	\label{tab:3D3}
\end{table} 

\subsection{Reference momentum dependence} 
\begin{figure}[h]
	\centering 
	\includegraphics[width=8cm]{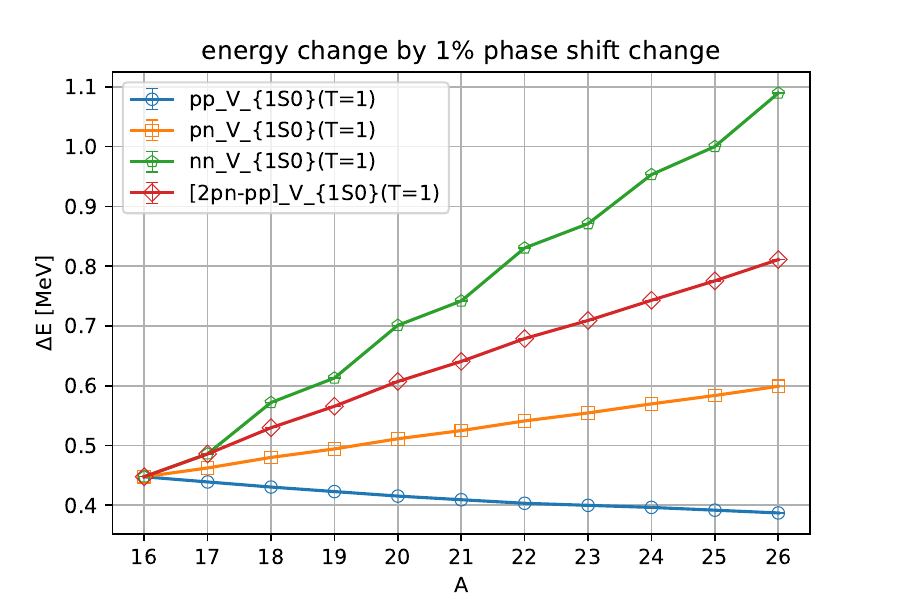} 
	\includegraphics[width=8cm]{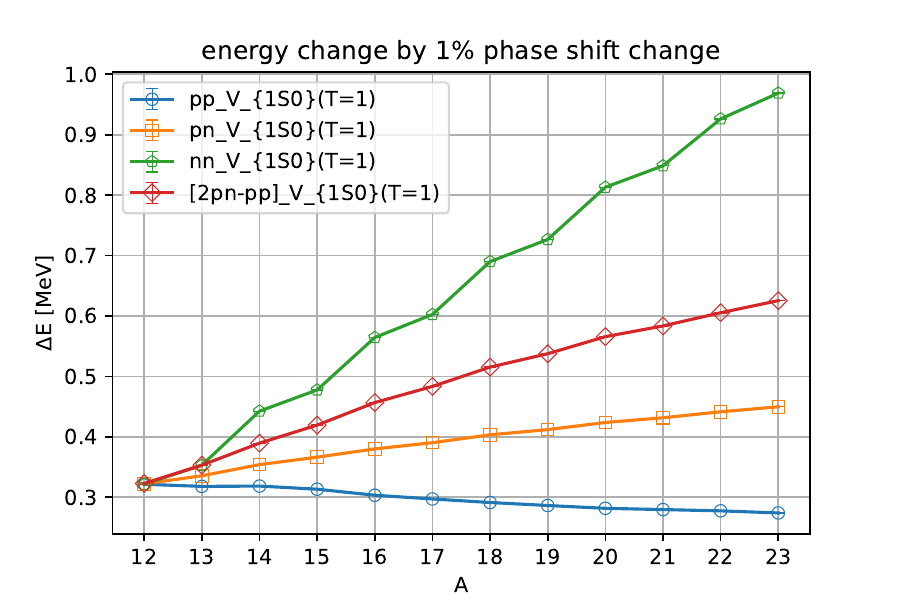} 
	\caption{Correlations for pp, pn, nn, and 2pn$-$pp in the $^1$S$_0$ channel.  The top panel shows the oxygen isotopes, and the bottom panel shows the carbon isotopes. reference momentum $p=100$ MeV.}
	\label{fig:1S0_100}
\end{figure}

\begin{figure}[h]
	\centering 
	\includegraphics[width=8cm]{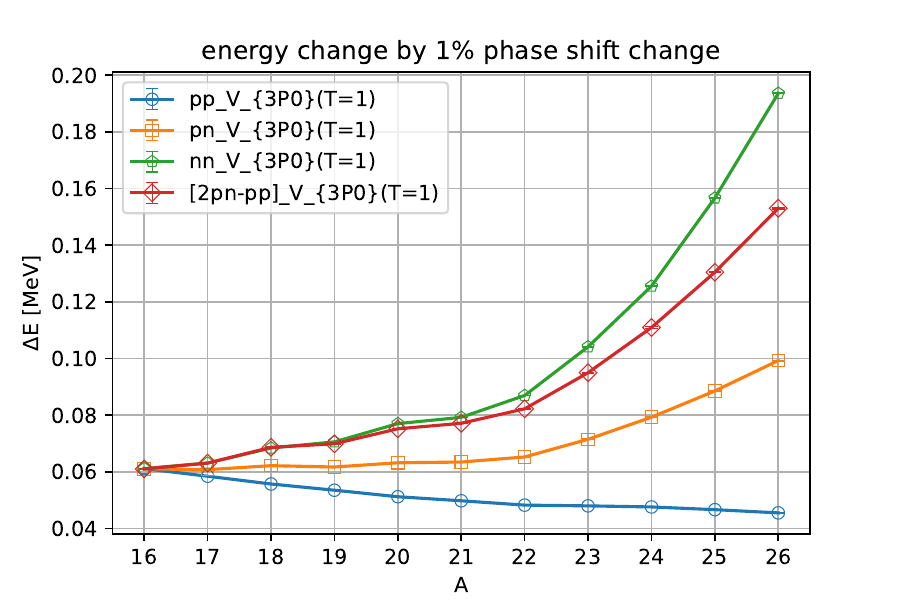} 
	\includegraphics[width=8cm]{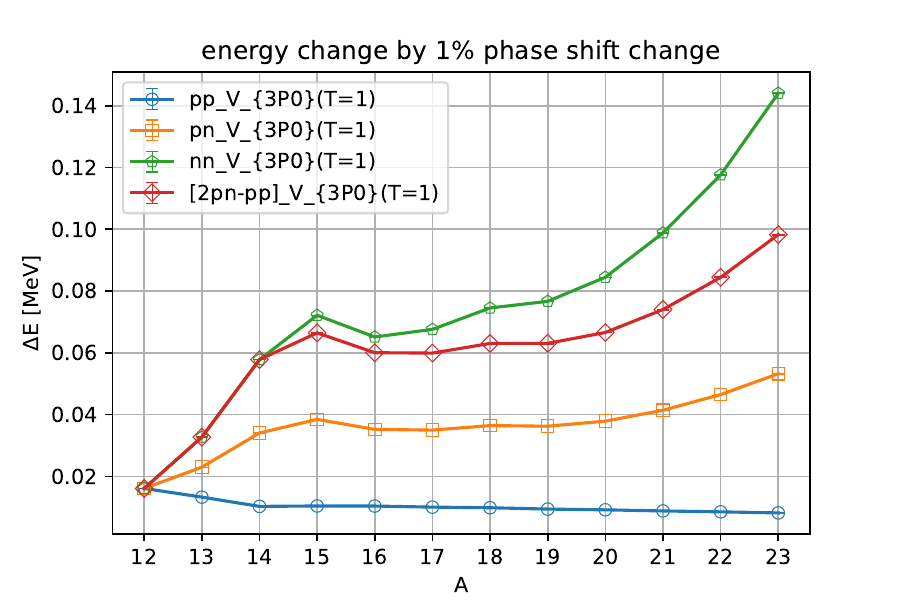} 
	\caption{Correlations for pp, pn, nn, and 2pn$-$pp in the $^3$P$_0$ channel.  The top panel shows the oxygen isotopes, and the bottom panel shows the carbon isotopes. reference momentum $p=100$ MeV.}
	\label{fig:3P0_100}
\end{figure}

\begin{figure}[h]
	\centering 
	\includegraphics[width=8cm]{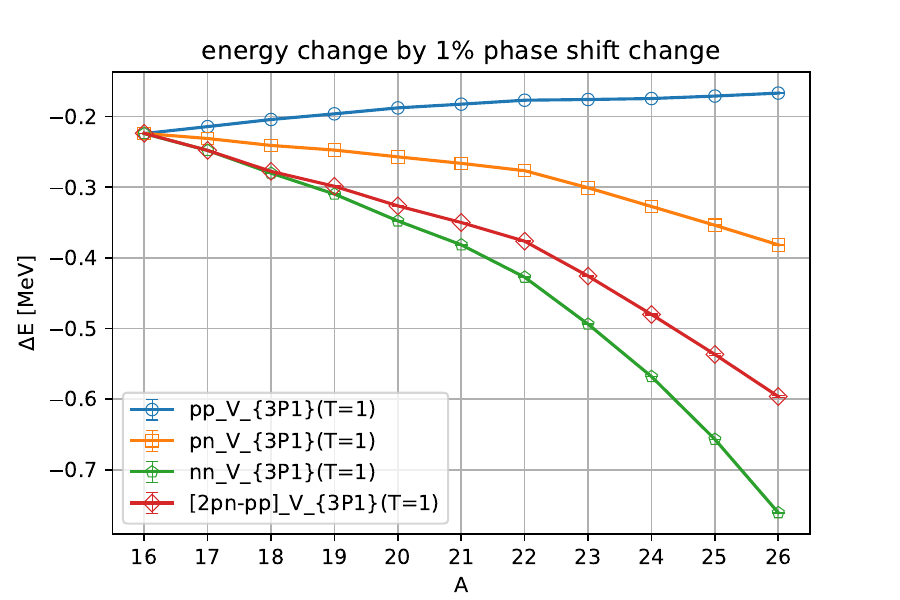}
	\includegraphics[width=8cm]{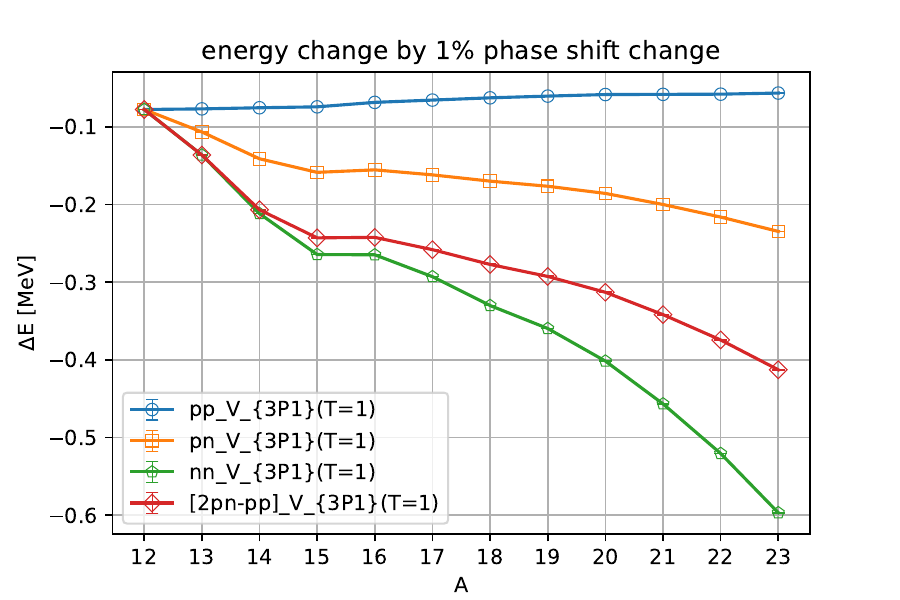}
	\caption{Correlations for pp, pn, nn, and 2pn$-$pp in the $^3$P$_1$ channel.  The left panel shows the oxygen isotopes, and the right panel shows the carbon isotopes. reference momentum $p=100$ MeV.}
	\label{fig:3P1_100}
\end{figure}

\begin{figure}[h]
	\centering 
	\includegraphics[width=8cm]{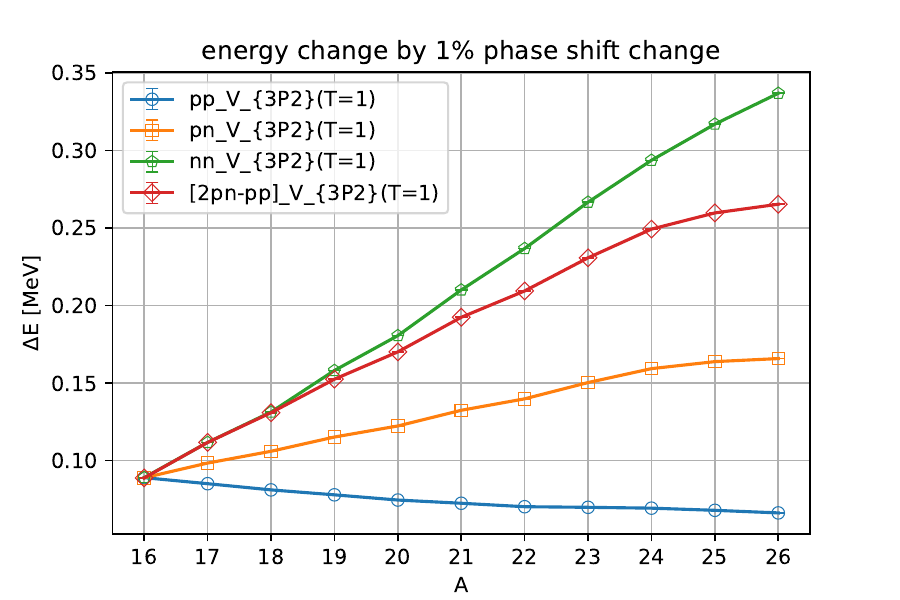}
	\includegraphics[width=8cm]{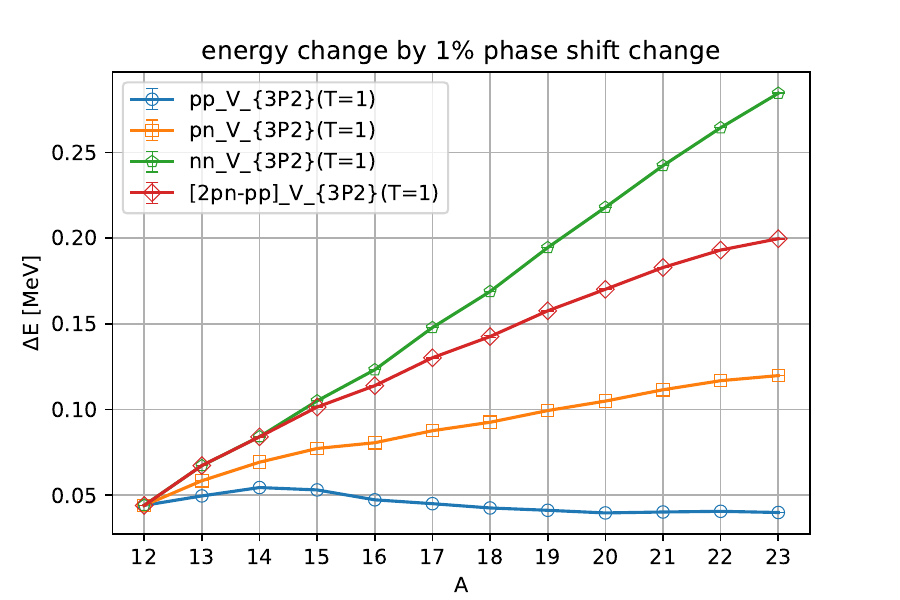}
	\caption{Correlations for pp, pn, nn, and 2pn$-$pp in the $^3$P$_2$ channel.  The left panel shows the oxygen isotopes, and the right panel shows the carbon isotopes. reference momentum $p=100$ MeV.}
	\label{fig:3P2_100}
\end{figure}

\begin{figure}[h]
	\centering 
	\includegraphics[width=8cm]{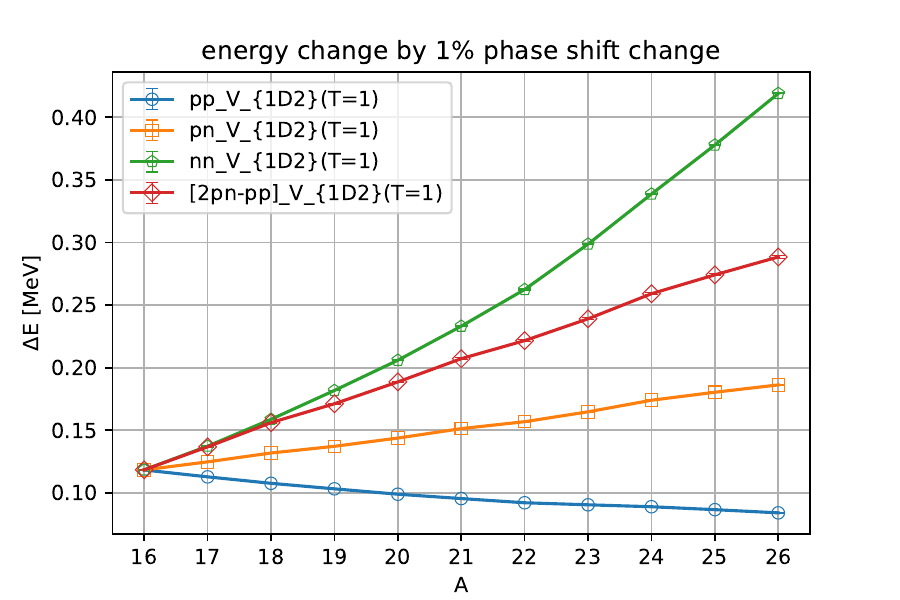}
	\includegraphics[width=8cm]{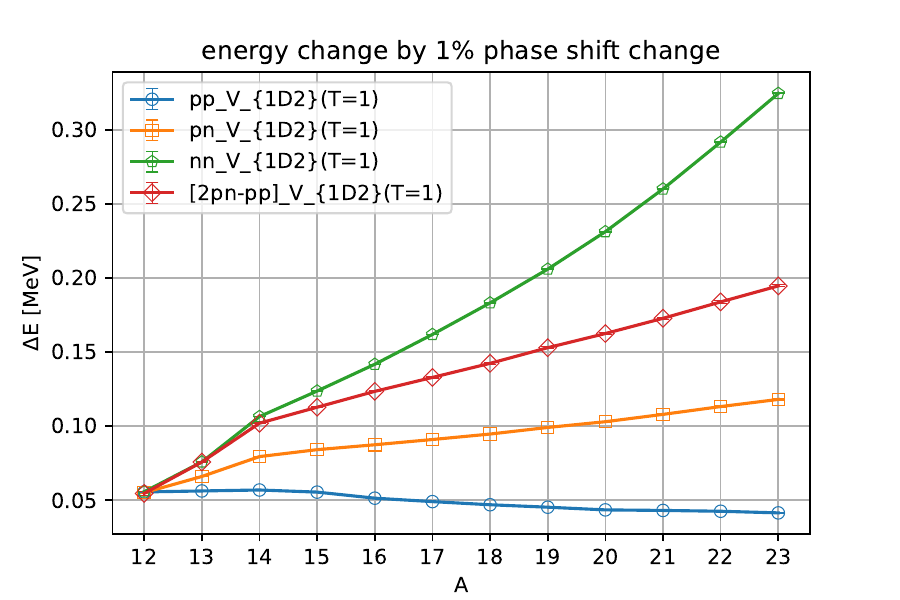}
	\caption{Correlations for pp, pn, nn, and 2pn$-$pp in the $^1$D$_2$ channel.  The left panel shows the oxygen isotopes, and the right panel shows the carbon isotopes. reference momentum $p=100$ MeV.}
	\label{fig:1D2_100}
\end{figure}

\begin{figure}[h]
	\centering 
	\includegraphics[width=8cm]{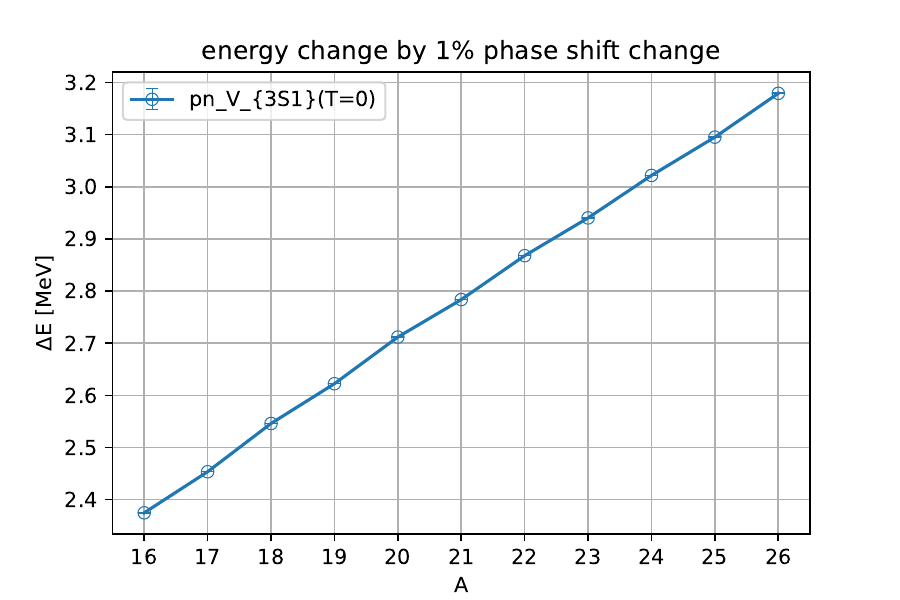}
	\includegraphics[width=8cm]{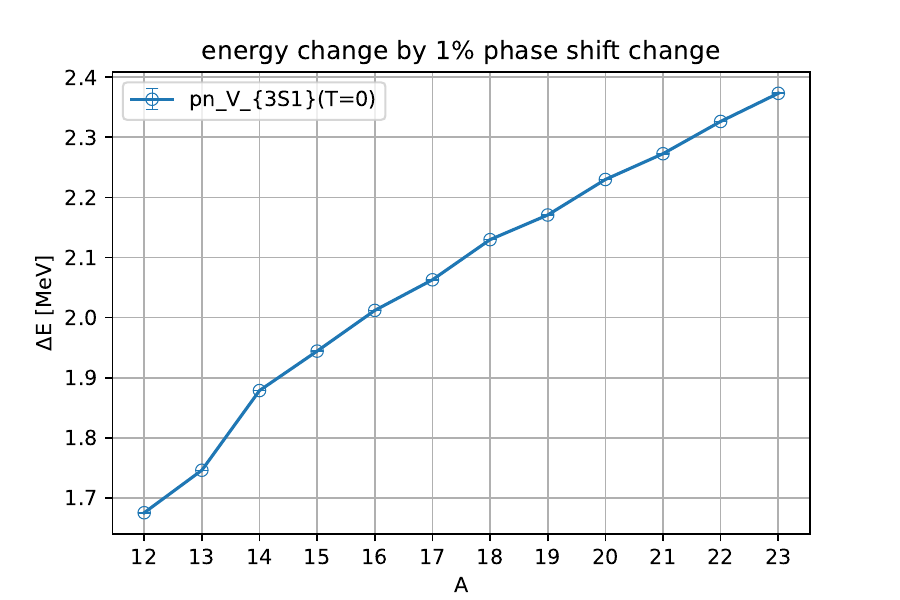}
	\caption{Correlations for pn in the $^3$S$_1$ channel.  The left panel shows the oxygen isotopes, and the right panel shows the carbon isotopes. reference momentum $p=100$ MeV.}
	\label{fig:3S1_100}
\end{figure}

\begin{figure}[h]
	\centering 
	\includegraphics[width=8cm]{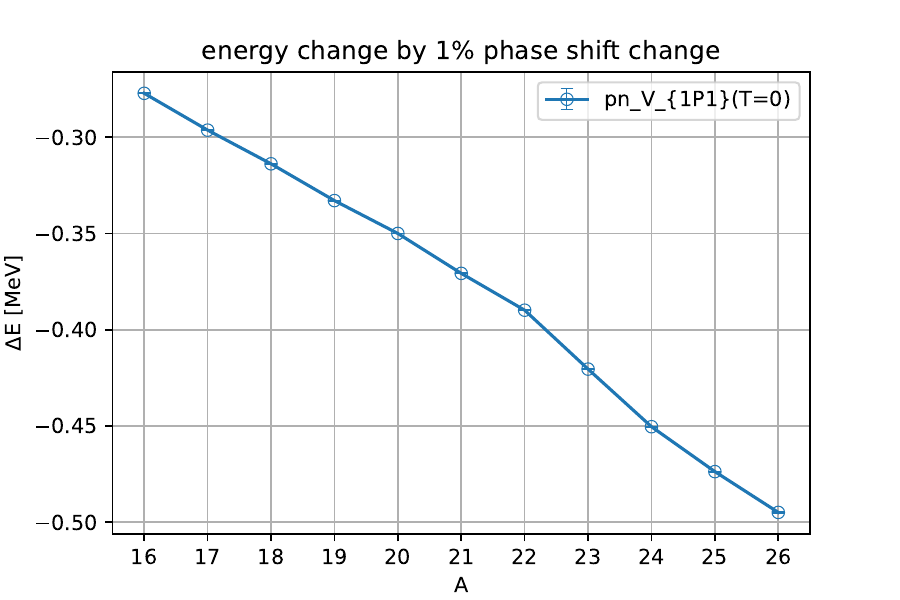}
	\includegraphics[width=8cm]{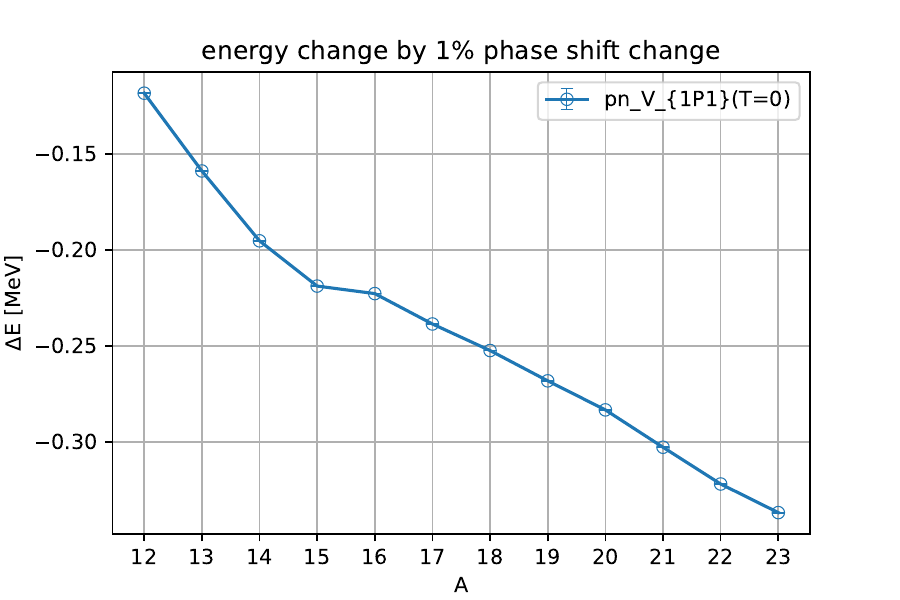}
	\caption{Correlations for pn in the $^1$P$_1$ channel.  The left panel shows the oxygen isotopes, and the right panel shows the carbon isotopes. reference momentum $p=100$ MeV.}
	\label{fig:1P1_100}
\end{figure}

\begin{figure}[h]
	\centering 
	\includegraphics[width=8cm]{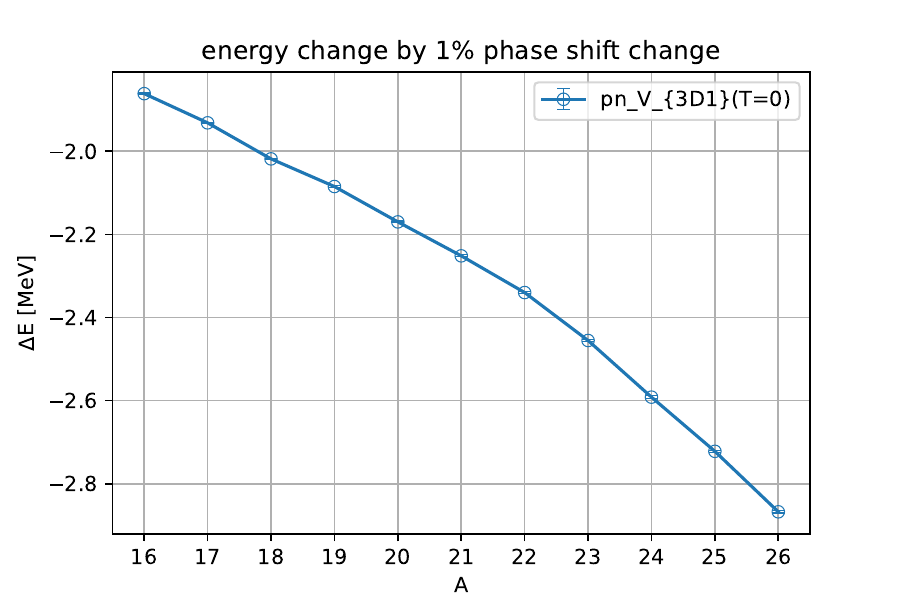}
	\includegraphics[width=8cm]{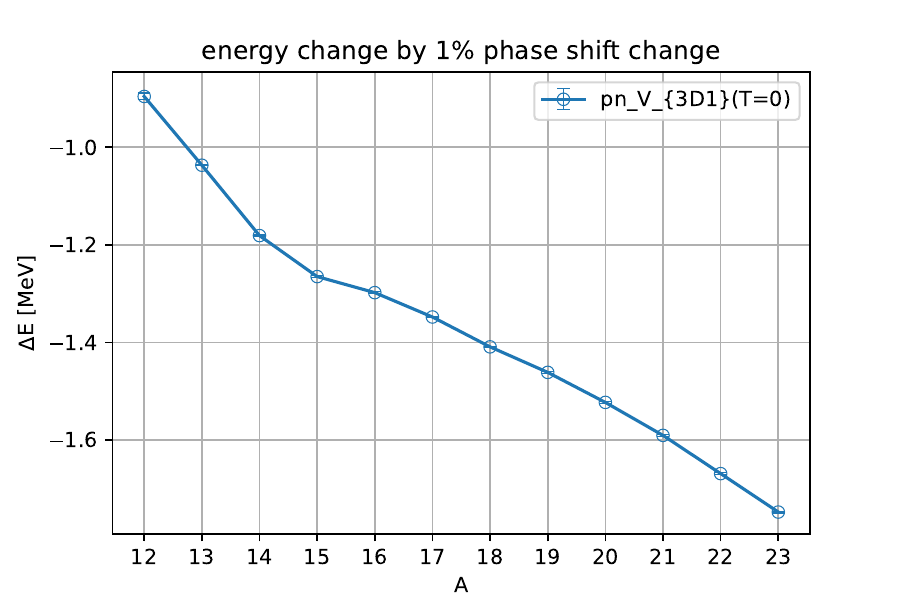}
	\caption{Correlations for pn in the $^3$D$_1$ channel.  The left panel shows the oxygen isotopes, and the right panel shows the carbon isotopes. reference momentum $p=100$ MeV.}
	\label{fig:3D1_100}
\end{figure}

\begin{figure}[h]
	\centering 
	\includegraphics[width=8cm]{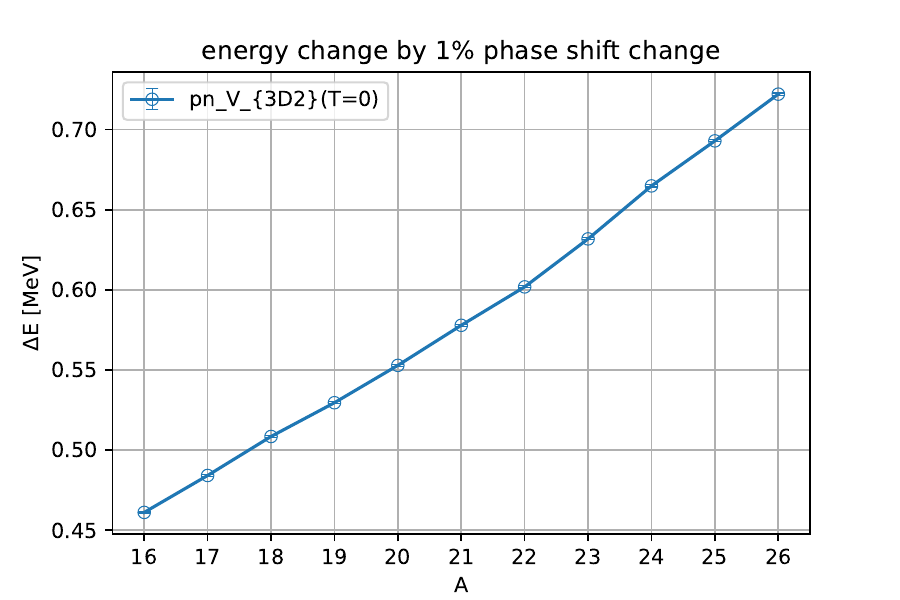}
	\includegraphics[width=8cm]{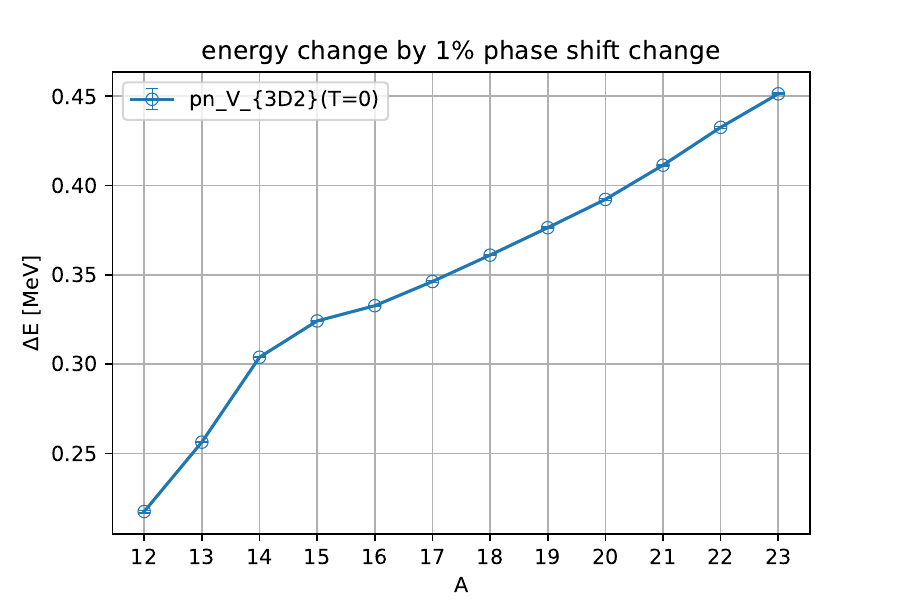}
	\caption{Correlations for pn in the $^3$D$_2$ channel.  The left panel shows the oxygen isotopes, and the right panel shows the carbon isotopes. reference momentum $p=100$ MeV.}
	\label{fig:3D2_100}
\end{figure}

\begin{figure}[h]
	\centering 
	\includegraphics[width=8cm]{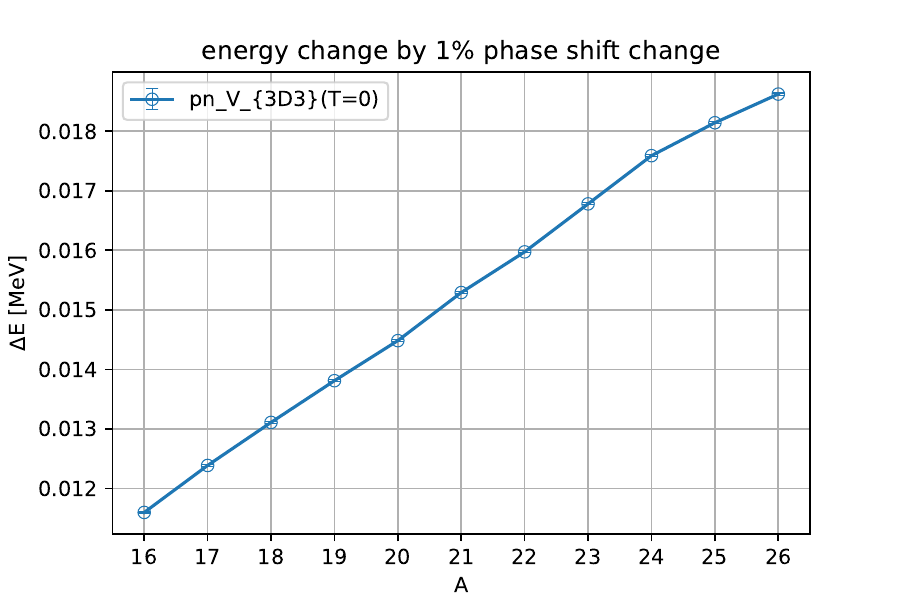}
	\includegraphics[width=8cm]{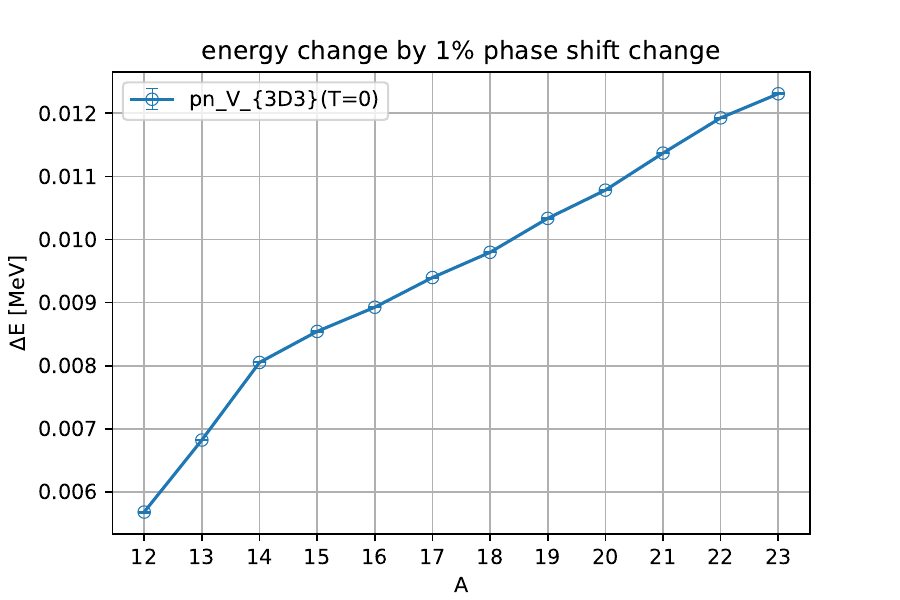}
	\caption{Correlations for pn in the $^3$D$_3$ channel.  The left panel shows the oxygen isotopes, and the right panel shows the carbon isotopes. reference momentum $p=100$ MeV.}
	\label{fig:3D3_100}
\end{figure}

In the main text, we consider a short range
two-nucleon interaction operator that, when added
to the full Hamiltonian, produces a 1\% reduction in
the scattering phase shift at relative momentum $p = 150$ MeV.
We have chosen operators with the same form as those in our Hamiltonian, 
corresponding to the lowest order short-range operators 
for our chosen nucleon-nucleon interaction channel.
This is a good starting point for exploring two-body correlations 
between nucleons that are relevant for nuclear binding.
On the other hand, we choose the reference momentum $p = 150$ MeV, 
because it is close to the momentum resolution for spatial lattice spacing 
$a=1.32$ fm in our calculation. However, the choice of reference momentum 
is not unique. 
We have also  repeated our analysis with another reference momentum, $p = 100$ MeV.
As expected, we obtain exactly the same results with 
only some overall renormalization of the expectation values.
The conventions are the same as those used for the  $p=150$ MeV case,
and the corresponding results are given in Figures \ref{fig:1S0_100} to \ref{fig:3D3_100}
and Tables \ref{tab:1S0_100} to \ref{tab:3D3_100}. 

\begin{table}
	\centering
	\begin{tabular}{|r|r|r|r|r|}
		\hline
		Channel & Momentum (MeV) & Phase Shift (deg) & New Phase Shift (deg) & Change \\
		\hline
		\hline
		$V_{\rm 1S0}$   &     50  & 6.356E+01  &  6.256E+01   & -1.56\% \\
		$V_{\rm 1S0}$   &    100  & 5.317E+01  &  5.264E+01   & -1.00\% \\
		$V_{\rm 1S0}$   &    150  & 4.136E+01  &  4.099E+01   & -0.88\% \\
		$V_{\rm 3P0}$   &     50  & 1.823E+00  &  1.813E+00   & -0.54\% \\
		$V_{\rm 3P0}$   &    100  & 7.039E+00  &  6.969E+00   & -0.99\% \\
		$V_{\rm 3P0}$   &    150  & 1.003E+01  &  9.863E+00   & -1.72\% \\
		$V_{\rm 3P1}$   &     50  & -1.120E+00 &   -1.113E+00 &   -0.62\% \\
		$V_{\rm 3P1}$   &    100  & -4.475E+00 &   -4.430E+00 &   -1.01\% \\
		$V_{\rm 3P1}$   &    150  & -8.091E+00 &   -7.976E+00 &   -1.43\% \\
		$V_{\rm 3P2}$   &     50  & 2.557E-01  &  2.533E-01   & -0.93\% \\
		$V_{\rm 3P2}$   &    100  & 2.023E+00  &  2.003E+00   & -1.00\% \\
		$V_{\rm 3P2}$   &    150  & 5.723E+00  &  5.656E+00   & -1.17\% \\
		$V_{\rm 1D2}$   &     50  & 5.333E-02  &  5.317E-02   & -0.31\% \\
		$V_{\rm 1D2}$   &    100  & 5.255E-01  &  5.203E-01   & -0.99\% \\
		$V_{\rm 1D2}$   &    150  & 1.414E+00  &  1.377E+00   & -2.65\% \\
		$V_{\rm 3S1}$   &     50  & 1.168E+02  &  1.159E+02   & -0.78\% \\
		$V_{\rm 3S1}$   &    100  & 8.462E+01  &  8.378E+01   & -1.00\% \\
		$V_{\rm 3S1}$   &    150  & 6.378E+01  &  6.300E+01   & -1.22\% \\
		$V_{\rm 1P1}$   &     50  & -1.683E+00 &   -1.674E+00 &   -0.56\% \\
		$V_{\rm 1P1}$   &    100  & -5.905E+00 &   -5.846E+00 &   -1.01\% \\
		$V_{\rm 1P1}$   &    150  & -9.738E+00 &   -9.590E+00 &   -1.52\% \\
		$V_{\rm 3D1}$   &     50  & -2.191E-01 &   -2.174E-01 &   -0.77\% \\
		$V_{\rm 3D1}$   &    100  & -2.311E+00 &   -2.288E+00 &   -1.00\% \\
		$V_{\rm 3D1}$   &    150  & -6.207E+00 &   -6.161E+00 &   -0.75\% \\
		$V_{\rm 3D2}$   &     50  & 2.596E-01  &  2.589E-01   & -0.29\% \\
		$V_{\rm 3D2}$   &    100  & 2.859E+00  &  2.830E+00   & -1.00\% \\
		$V_{\rm 3D2}$   &    150  & 7.581E+00  &  7.361E+00   & -2.89\% \\
		$V_{\rm 3D3}$   &     50  & 6.407E-03  &  6.397E-03   & -0.16\% \\
		$V_{\rm 3D3}$   &    100  & 3.245E-02  &  3.213E-02   & -1.00\% \\
		$V_{\rm 3D3}$   &    150  & 3.070E-01  &  3.047E-01   & -0.75\% \\
		\hline
	\end{tabular}
	\caption{Scattering phase shifts and phase shift changes produced by the two-nucleon operator perturbations at reference momentum $p=100$ MeV.}
	\label{tab:phase_shifts_100}
\end{table}

\begin{table}
	\centering
	\begin{tabular}{|r|r|r|r|r|}
		\hline
		Nucleus	&	pp (MeV)	&	pn (MeV)	&	nn (MeV)	&	2pn$-$pp (MeV)	\\
		\hline
		\hline
		$^{16}$O & 0.4476(1) & 0.4476(1) & 0.4477(1) & 0.4477(1) \\
		$^{17}$O & 0.4390(1) & 0.4624(1) & 0.4861(1) & 0.4858(1) \\
		$^{18}$O & 0.4304(1) & 0.4800(1) & 0.5719(1) & 0.5296(2) \\
		$^{19}$O & 0.4229(1) & 0.4942(1) & 0.6129(1) & 0.5656(2) \\
		$^{20}$O & 0.4153(1) & 0.5112(1) & 0.7011(1) & 0.6071(2) \\
		$^{21}$O & 0.4092(1) & 0.5249(1) & 0.7417(1) & 0.6406(2) \\
		$^{22}$O & 0.4033(1) & 0.5411(2) & 0.8305(1) & 0.6789(4) \\
		$^{23}$O & 0.3998(1) & 0.5545(2) & 0.8711(1) & 0.7091(4) \\
		$^{24}$O & 0.3964(1) & 0.5696(2) & 0.9532(1) & 0.7429(3) \\
		$^{25}$O & 0.3917(1) & 0.5836(2) & 1.0002(1) & 0.7755(3) \\
		$^{26}$O & 0.3871(1) & 0.5991(2) & 1.0895(2) & 0.8110(4) \\  
		\hline
	\end{tabular}
	\quad
	\quad
	\begin{tabular}{|r|r|r|r|r|}
		\hline
		Nucleus	&	pp (MeV)	&	pn (MeV)	&	nn (MeV)	&	2pn$-$pp (MeV)	\\
		\hline
		\hline
		$^{12}$C  & 0.3216(1) & 0.3223(6) & 0.3216(1) & 0.3230(11) \\
		$^{13}$C  & 0.3181(1) & 0.3357(1) & 0.3532(1) & 0.3532(1) \\
		$^{14}$C  & 0.3186(1) & 0.3541(1) & 0.4426(1) & 0.3896(1) \\
		$^{15}$C  & 0.3134(1) & 0.3665(1) & 0.4777(1) & 0.4195(1) \\
		$^{16}$C  & 0.3035(1) & 0.3801(1) & 0.5644(1) & 0.4568(2) \\
		$^{17}$C  & 0.2973(1) & 0.3904(1) & 0.6026(1) & 0.4835(1) \\
		$^{18}$C  & 0.2914(1) & 0.4034(2) & 0.6901(1) & 0.5155(3) \\
		$^{19}$C  & 0.2864(1) & 0.4120(1) & 0.7267(1) & 0.5376(2) \\
		$^{20}$C  & 0.2818(1) & 0.4238(2) & 0.8131(1) & 0.5659(3) \\
		$^{21}$C  & 0.2797(1) & 0.4316(2) & 0.8487(1) & 0.5836(3) \\
		$^{22}$C  & 0.2776(1) & 0.4415(2) & 0.9262(2) & 0.6055(4) \\
		$^{23}$C  & 0.2741(1) & 0.4498(2) & 0.9688(1) & 0.6254(3) \\     
		\hline
	\end{tabular}
	\caption{$^1$S$_0$ correlations for the oxygen and carbon isotopes with reference momentum $p=100$ MeV.}
	\label{tab:1S0_100}
\end{table}

\begin{table}
	\centering
	\begin{tabular}{|r|r|r|r|r|}
		\hline
		Nucleus	&	pp (MeV)	&	pn (MeV)	&	nn (MeV)	&	2pn$-$pp (MeV)	\\
		\hline
		\hline
		$^{16}$O & 0.0612(1) & 0.0611(1) & 0.0612(1) & 0.0611(1) \\
		$^{17}$O & 0.0585(1) & 0.0608(1) & 0.0631(1) & 0.0631(1) \\
		$^{18}$O & 0.0557(1) & 0.0622(1) & 0.0684(1) & 0.0687(1) \\
		$^{19}$O & 0.0535(1) & 0.0617(1) & 0.0706(1) & 0.0699(1) \\
		$^{20}$O & 0.0512(1) & 0.0632(1) & 0.0770(1) & 0.0752(1) \\
		$^{21}$O & 0.0498(1) & 0.0635(1) & 0.0793(1) & 0.0772(1) \\
		$^{22}$O & 0.0483(1) & 0.0653(1) & 0.0870(1) & 0.0823(1) \\
		$^{23}$O & 0.0480(1) & 0.0715(1) & 0.1042(1) & 0.0950(1) \\
		$^{24}$O & 0.0476(1) & 0.0793(1) & 0.1256(1) & 0.1110(2) \\
		$^{25}$O & 0.0467(1) & 0.0886(1) & 0.1567(1) & 0.1305(2) \\
		$^{26}$O & 0.0455(1) & 0.0993(1) & 0.1936(1) & 0.1530(2) \\  
		\hline
	\end{tabular}
	\quad
	\quad
	\begin{tabular}{|r|r|r|r|r|}
		\hline
		Nucleus	&	pp (MeV)	&	pn (MeV)	&	nn (MeV)	&	2pn$-$pp (MeV)	\\
		\hline
		\hline
		$^{12}$C & 0.0161(1) & 0.0161(1) & 0.0161(1) & 0.0161(1) \\
		$^{13}$C & 0.0133(1) & 0.0230(1) & 0.0327(1) & 0.0327(1) \\
		$^{14}$C & 0.0103(1) & 0.0340(1) & 0.0578(1) & 0.0578(1) \\
		$^{15}$C & 0.0104(1) & 0.0384(1) & 0.0721(1) & 0.0665(1) \\
		$^{16}$C & 0.0104(1) & 0.0352(1) & 0.0651(1) & 0.0600(1) \\
		$^{17}$C & 0.0100(1) & 0.0350(1) & 0.0676(1) & 0.0599(1) \\
		$^{18}$C & 0.0098(1) & 0.0364(1) & 0.0745(1) & 0.0630(1) \\
		$^{19}$C & 0.0094(1) & 0.0362(1) & 0.0767(1) & 0.0631(2) \\
		$^{20}$C & 0.0092(1) & 0.0379(1) & 0.0845(1) & 0.0666(1) \\
		$^{21}$C & 0.0088(1) & 0.0414(1) & 0.0988(1) & 0.0740(1) \\
		$^{22}$C & 0.0085(1) & 0.0465(1) & 0.1176(1) & 0.0844(2) \\
		$^{23}$C & 0.0082(1) & 0.0532(1) & 0.1440(1) & 0.0983(1) \\ 
		\hline
	\end{tabular}
	\caption{$^3$P$_0$ correlations for the oxygen and carbon isotopes
		with reference momentum $p=100$ MeV.}
	\label{tab:3P0_100}
\end{table}

\begin{table}
	\centering
	\begin{tabular}{|r|r|r|r|r|}
		\hline
		Nucleus	&	pp (MeV)	&	pn (MeV)	&	nn (MeV)	&	2pn$-$pp (MeV)	\\
		\hline
		\hline
		$^{16}$O   & -0.2242(1) & -0.2240(2) & -0.2243(1) & -0.2238(3) \\
		$^{17}$O   & -0.2144(1) & -0.2312(2) & -0.2481(1) & -0.2480(3) \\
		$^{18}$O   & -0.2043(1) & -0.2410(3) & -0.2802(1) & -0.2777(4) \\
		$^{19}$O   & -0.1963(1) & -0.2476(3) & -0.3098(2) & -0.2989(4) \\
		$^{20}$O   & -0.1878(1) & -0.2572(3) & -0.3478(2) & -0.3265(5) \\
		$^{21}$O   & -0.1826(1) & -0.2664(3) & -0.3817(2) & -0.3501(5) \\
		$^{22}$O   & -0.1770(1) & -0.2768(4) & -0.4275(3) & -0.3765(8) \\
		$^{23}$O   & -0.1760(1) & -0.3009(4) & -0.4940(2) & -0.4259(6) \\
		$^{24}$O   & -0.1746(1) & -0.3274(4) & -0.5682(3) & -0.4802(7) \\
		$^{25}$O   & -0.1711(1) & -0.3540(5) & -0.6569(3) & -0.5368(8) \\
		$^{26}$O   & -0.1669(1) & -0.3817(5) & -0.7608(4) & -0.5964(9) \\
		\hline
	\end{tabular}
	\quad
	\quad
	\begin{tabular}{|r|r|r|r|r|}
		\hline
		Nucleus	&	pp (MeV)	&	pn (MeV)	&	nn (MeV)	&	2pn$-$pp (MeV)	\\
		\hline
		\hline
		$^{12}$C  & -0.0778(1) & -0.0778(1) & -0.0779(1) & -0.0778(1) \\
		$^{13}$C  & -0.0770(1) & -0.1067(1) & -0.1364(1) & -0.1364(2) \\
		$^{14}$C  & -0.0756(1) & -0.1413(1) & -0.2118(1) & -0.2069(2) \\
		$^{15}$C  & -0.0744(1) & -0.1587(2) & -0.2645(1) & -0.2430(3) \\
		$^{16}$C  & -0.0687(1) & -0.1555(1) & -0.2648(1) & -0.2424(3) \\
		$^{17}$C  & -0.0657(1) & -0.1620(2) & -0.2930(1) & -0.2582(3) \\
		$^{18}$C  & -0.0628(1) & -0.1700(3) & -0.3301(3) & -0.2772(6) \\
		$^{19}$C  & -0.0606(1) & -0.1766(3) & -0.3597(2) & -0.2927(7) \\
		$^{20}$C  & -0.0586(1) & -0.1858(2) & -0.4016(2) & -0.3131(4) \\
		$^{21}$C  & -0.0583(1) & -0.2001(3) & -0.4567(2) & -0.3418(5) \\
		$^{22}$C  & -0.0581(1) & -0.2163(3) & -0.5205(3) & -0.3744(6) \\
		$^{23}$C  & -0.0567(1) & -0.2347(3) & -0.5968(3) & -0.4128(6) \\
		\hline
	\end{tabular}
	\caption{$^3$P$_1$ correlations for the oxygen and carbon isotopes
		with reference momentum $p=100$ MeV.}
	\label{tab:3P1_100}
\end{table}

\begin{table}
	\centering
	\begin{tabular}{|r|r|r|r|r|}
		\hline
		Nucleus	&	pp (MeV)	&	pn (MeV)	&	nn (MeV)	&	2pn$-$pp (MeV)	\\
		\hline
		\hline
		$^{16}$O  & 0.0890(1) & 0.0890(1) & 0.0891(1) & 0.0889(1) \\
		$^{17}$O  & 0.0851(1) & 0.0984(1) & 0.1118(1) & 0.1118(1) \\
		$^{18}$O  & 0.0811(1) & 0.1060(1) & 0.1313(1) & 0.1309(2) \\
		$^{19}$O  & 0.0780(1) & 0.1152(1) & 0.1581(1) & 0.1525(2) \\
		$^{20}$O  & 0.0746(1) & 0.1224(1) & 0.1806(1) & 0.1701(2) \\
		$^{21}$O  & 0.0725(1) & 0.1325(1) & 0.2101(1) & 0.1924(2) \\
		$^{22}$O  & 0.0703(1) & 0.1399(2) & 0.2368(2) & 0.2094(4) \\
		$^{23}$O  & 0.0699(1) & 0.1504(1) & 0.2667(1) & 0.2308(3) \\
		$^{24}$O  & 0.0694(1) & 0.1593(2) & 0.2937(1) & 0.2493(3) \\
		$^{25}$O  & 0.0680(1) & 0.1638(2) & 0.3170(1) & 0.2597(3) \\
		$^{26}$O  & 0.0663(1) & 0.1658(2) & 0.3369(1) & 0.2653(4) \\
		\hline
	\end{tabular}
	\quad
	\quad
	\begin{tabular}{|r|r|r|r|r|}
		\hline
		Nucleus	&	pp (MeV)	&	pn (MeV)	&	nn (MeV)	&	2pn$-$pp (MeV)	\\
		\hline
		\hline
		$^{12}$C  & 0.0441(1) & 0.0440(1) & 0.0441(1) & 0.0440(1) \\
		$^{13}$C  & 0.0496(1) & 0.0584(1) & 0.0673(1) & 0.0673(1) \\
		$^{14}$C  & 0.0545(1) & 0.0693(1) & 0.0841(1) & 0.0840(1) \\
		$^{15}$C  & 0.0531(1) & 0.0772(1) & 0.1050(1) & 0.1014(1) \\
		$^{16}$C  & 0.0473(1) & 0.0806(1) & 0.1232(1) & 0.1139(1) \\
		$^{17}$C  & 0.0450(1) & 0.0876(1) & 0.1478(1) & 0.1302(1) \\
		$^{18}$C  & 0.0425(1) & 0.0925(1) & 0.1688(1) & 0.1426(3) \\
		$^{19}$C  & 0.0412(1) & 0.0994(1) & 0.1944(1) & 0.1576(2) \\
		$^{20}$C  & 0.0396(1) & 0.1049(1) & 0.2179(1) & 0.1701(2) \\
		$^{21}$C  & 0.0402(1) & 0.1115(1) & 0.2422(1) & 0.1828(2) \\
		$^{22}$C  & 0.0406(1) & 0.1168(1) & 0.2644(1) & 0.1930(2) \\
		$^{23}$C  & 0.0399(1) & 0.1198(1) & 0.2845(1) & 0.1996(3) \\
		\hline
	\end{tabular}
	\caption{$^3$P$_2$ correlations for the oxygen and carbon isotopes
		with reference momentum $p=100$ MeV.}
	\label{tab:3P2_100}
\end{table}

\begin{table}
	\centering
	\begin{tabular}{|r|r|r|r|r|}
		\hline
		Nucleus	&	pp (MeV)	&	pn (MeV)	&	nn (MeV)	&	2pn$-$pp (MeV)	\\
		\hline
		\hline
		$^{16}$O  & 0.1183(1) & 0.1184(2) & 0.1184(1) & 0.1185(3) \\
		$^{17}$O  & 0.1129(1) & 0.1248(2) & 0.1373(1) & 0.1368(4) \\
		$^{18}$O  & 0.1077(1) & 0.1319(2) & 0.1588(1) & 0.1561(4) \\
		$^{19}$O  & 0.1033(1) & 0.1372(2) & 0.1818(1) & 0.1712(4) \\
		$^{20}$O  & 0.0990(1) & 0.1438(3) & 0.2059(1) & 0.1887(6) \\
		$^{21}$O  & 0.0955(1) & 0.1514(3) & 0.2331(2) & 0.2072(5) \\
		$^{22}$O  & 0.0922(1) & 0.1569(5) & 0.2624(4) & 0.2216(9) \\
		$^{23}$O  & 0.0905(1) & 0.1648(4) & 0.2987(2) & 0.2390(7) \\
		$^{24}$O  & 0.0890(1) & 0.1740(4) & 0.3386(2) & 0.2591(7) \\
		$^{25}$O  & 0.0867(1) & 0.1804(4) & 0.3778(3) & 0.2741(7) \\
		$^{26}$O  & 0.0841(1) & 0.1862(5) & 0.4189(3) & 0.2884(8) \\
		\hline
	\end{tabular}
	\quad
	\quad
	\begin{tabular}{|r|r|r|r|r|}
		\hline
		Nucleus	&	pp (MeV)	&	pn (MeV)	&	nn (MeV)	&	2pn$-$pp (MeV)	\\
		\hline
		\hline
		$^{12}$C  & 0.0555(1) & 0.0550(6) & 0.0556(1) & 0.0545(13) \\
		$^{13}$C  & 0.0562(1) & 0.0659(1) & 0.0758(1) & 0.0757(2) \\
		$^{14}$C  & 0.0568(1) & 0.0794(1) & 0.1064(1) & 0.1019(2) \\
		$^{15}$C  & 0.0554(1) & 0.0841(2) & 0.1236(1) & 0.1127(4) \\
		$^{16}$C  & 0.0513(1) & 0.0874(2) & 0.1417(1) & 0.1234(3) \\
		$^{17}$C  & 0.0490(1) & 0.0909(2) & 0.1618(1) & 0.1328(4) \\
		$^{18}$C  & 0.0469(1) & 0.0946(2) & 0.1831(1) & 0.1423(4) \\
		$^{19}$C  & 0.0452(1) & 0.0990(2) & 0.2059(1) & 0.1529(4) \\
		$^{20}$C  & 0.0434(1) & 0.1029(3) & 0.2312(2) & 0.1625(6) \\
		$^{21}$C  & 0.0430(1) & 0.1079(3) & 0.2600(2) & 0.1727(5) \\
		$^{22}$C  & 0.0425(1) & 0.1131(5) & 0.2917(2) & 0.1838(9) \\
		$^{23}$C  & 0.0414(1) & 0.1180(3) & 0.3245(2) & 0.1946(7) \\
		\hline
	\end{tabular}
	\caption{$^1$D$_2$ correlations for the oxygen and carbon isotopes
		with reference momentum $p=100$ MeV.}
	\label{tab:1D2_100}
\end{table}

\begin{table}
	\centering
	\begin{tabular}{|r|r|}
		\hline
		Nucleus	&	pn (MeV)	\\
		\hline
		\hline
		$^{16}$O  & 2.3748(3) \\
		$^{17}$O  & 2.4535(4) \\
		$^{18}$O  & 2.5460(4) \\
		$^{19}$O  & 2.6223(4) \\
		$^{20}$O  & 2.7118(4) \\
		$^{21}$O  & 2.7836(5) \\
		$^{22}$O  & 2.8678(5) \\
		$^{23}$O  & 2.9402(5) \\
		$^{24}$O  & 3.0217(5) \\
		$^{25}$O  & 3.0951(5) \\
		$^{26}$O  & 3.1791(6) \\
		\hline
	\end{tabular}
	\quad
	\quad
	\begin{tabular}{|r|r|}
		\hline
		Nucleus	&	pn (MeV)	\\
		\hline
		\hline
		$^{12}$C  & 1.6754(4) \\
		$^{13}$C  & 1.7461(2) \\
		$^{14}$C  & 1.8786(3) \\
		$^{15}$C  & 1.9442(3) \\
		$^{16}$C  & 2.0119(3) \\
		$^{17}$C  & 2.0628(3) \\
		$^{18}$C  & 2.1296(3) \\
		$^{19}$C  & 2.1706(4) \\
		$^{20}$C  & 2.2297(4) \\
		$^{21}$C  & 2.2725(4) \\
		$^{22}$C  & 2.3262(4) \\
		$^{23}$C  & 2.3730(5) \\
		\hline
	\end{tabular}
	\caption{$^3$S$_1$ correlations for the oxygen and carbon isotopes with reference momentum $p=100$ MeV}
	\label{tab:3S1_100}
\end{table} 

\begin{table}
	\centering
	\begin{tabular}{|r|r|}
		\hline
		Nucleus	&	pn (MeV)	\\
		\hline
		\hline
		$^{16}$O  & -0.2772(1) \\
		$^{17}$O  & -0.2964(1) \\
		$^{18}$O  & -0.3139(2) \\
		$^{19}$O  & -0.3330(2) \\
		$^{20}$O  & -0.3500(2) \\
		$^{21}$O  & -0.3708(2) \\
		$^{22}$O  & -0.3899(2) \\
		$^{23}$O  & -0.4205(2) \\
		$^{24}$O  & -0.4504(2) \\
		$^{25}$O  & -0.4737(2) \\
		$^{26}$O  & -0.4949(3) \\
		\hline
	\end{tabular}
	\quad
	\quad
	\begin{tabular}{|r|r|}
		\hline
		Nucleus	&	pn (MeV)	\\
		\hline
		\hline
		$^{12}$C  & -0.1184(1) \\
		$^{13}$C  & -0.1589(1) \\
		$^{14}$C  & -0.1953(1) \\
		$^{15}$C  & -0.2189(1) \\
		$^{16}$C  & -0.2228(1) \\
		$^{17}$C  & -0.2386(1) \\
		$^{18}$C  & -0.2524(1) \\
		$^{19}$C  & -0.2682(1) \\
		$^{20}$C  & -0.2833(1) \\
		$^{21}$C  & -0.3028(1) \\
		$^{22}$C  & -0.3219(1) \\
		$^{23}$C  & -0.3368(2) \\
		\hline
	\end{tabular}
	\caption{$^1$P$_1$ correlations for the oxygen and carbon isotopes with reference momentum $p=100$ MeV}
	\label{tab:1P1_100}
\end{table} 

\begin{table}
	\centering
	\begin{tabular}{|r|r|}
		\hline
		Nucleus	&	pn (MeV)	\\
		\hline
		\hline
		$^{16}$O  & -1.8615(15) \\
		$^{17}$O  & -1.9317(17) \\
		$^{18}$O  & -2.0184(19) \\
		$^{19}$O  & -2.0849(20) \\
		$^{20}$O  & -2.1700(21) \\
		$^{21}$O  & -2.2517(22) \\
		$^{22}$O  & -2.3400(23) \\
		$^{23}$O  & -2.4553(24) \\
		$^{24}$O  & -2.5915(28) \\
		$^{25}$O  & -2.7219(26) \\
		$^{26}$O  & -2.8674(28) \\
		\hline
	\end{tabular}
	\quad
	\quad
	\begin{tabular}{|r|r|}
		\hline
		Nucleus	&	pn (MeV)	\\
		\hline
		\hline
		$^{12}$C  & -0.8961(70) \\
		$^{13}$C  & -1.0370(12) \\
		$^{14}$C  & -1.1812(12) \\
		$^{15}$C  & -1.2652(15) \\
		$^{16}$C  & -1.2980(14) \\
		$^{17}$C  & -1.3480(14) \\
		$^{18}$C  & -1.4092(14) \\
		$^{19}$C  & -1.4614(17) \\
		$^{20}$C  & -1.5229(20) \\
		$^{21}$C  & -1.5906(18) \\
		$^{22}$C  & -1.6689(19) \\
		$^{23}$C  & -1.7476(19) \\
		\hline
	\end{tabular}
	\caption{$^3$D$_1$ correlations for the oxygen and carbon isotopes with reference momentum $p=100$ MeV}
	\label{tab:3D1_100}
\end{table} 

\begin{table}
	\centering
	\begin{tabular}{|r|r|}
		\hline
		Nucleus	&	pn (MeV)	\\
		\hline
		\hline
		$^{16}$O  & 0.4611(3) \\
		$^{17}$O  & 0.4841(4) \\
		$^{18}$O  & 0.5084(4) \\
		$^{19}$O  & 0.5295(4) \\
		$^{20}$O  & 0.5528(5) \\
		$^{21}$O  & 0.5778(5) \\
		$^{22}$O  & 0.6018(5) \\
		$^{23}$O  & 0.6318(5) \\
		$^{24}$O  & 0.6648(6) \\
		$^{25}$O  & 0.6930(6) \\
		$^{26}$O  & 0.7221(7) \\
		\hline
	\end{tabular}
	\quad
	\quad
	\begin{tabular}{|r|r|}
		\hline
		Nucleus	&	pn (MeV)	\\
		\hline
		\hline
		$^{12}$C  & 0.2174(8) \\
		$^{13}$C  & 0.2563(2) \\
		$^{14}$C  & 0.3038(2) \\
		$^{15}$C  & 0.3241(3) \\
		$^{16}$C  & 0.3327(3) \\
		$^{17}$C  & 0.3462(3) \\
		$^{18}$C  & 0.3610(3) \\
		$^{19}$C  & 0.3763(3) \\
		$^{20}$C  & 0.3922(4) \\
		$^{21}$C  & 0.4113(4) \\
		$^{22}$C  & 0.4325(4) \\
		$^{23}$C  & 0.4513(4) \\
		\hline
	\end{tabular}
	\caption{$^3$D$_2$ correlations for the oxygen and carbon isotopes with reference momentum $p=100$ MeV}
	\label{tab:3D2_100}
\end{table} 

\begin{table}
	\centering
	\begin{tabular}{|r|r|}
		\hline
		Nucleus	&	pn (MeV)	\\
		\hline
		\hline
		$^{16}$O  & 0.0116(1) \\
		$^{17}$O  & 0.0124(1) \\
		$^{18}$O  & 0.0131(1) \\
		$^{19}$O  & 0.0138(1) \\
		$^{20}$O  & 0.0145(1) \\
		$^{21}$O  & 0.0153(1) \\
		$^{22}$O  & 0.0160(1) \\
		$^{23}$O  & 0.0168(1) \\
		$^{24}$O  & 0.0176(1) \\
		$^{25}$O  & 0.0181(1) \\
		$^{26}$O  & 0.0186(1) \\
		\hline
	\end{tabular}
	\quad
	\quad
	\begin{tabular}{|r|r|}
		\hline
		Nucleus	&	pn (MeV)	\\
		\hline
		\hline
		$^{12}$C  & 0.0057(1) \\
		$^{13}$C  & 0.0068(1) \\
		$^{14}$C  & 0.0080(1) \\
		$^{15}$C  & 0.0085(1) \\
		$^{16}$C  & 0.0089(1) \\
		$^{17}$C  & 0.0094(1) \\
		$^{18}$C  & 0.0098(1) \\
		$^{19}$C  & 0.0103(1) \\
		$^{20}$C  & 0.0108(1) \\
		$^{21}$C  & 0.0114(1) \\
		$^{22}$C  & 0.0119(1) \\
		$^{23}$C  & 0.0123(1) \\
		\hline
	\end{tabular}
	\caption{$^3$D$_3$ correlations for the oxygen and carbon isotopes with reference momentum $p=100$ MeV}
	\label{tab:3D3_100}
\end{table}


\end{document}